\documentclass[11pt]{article}
\usepackage[T1]{fontenc}

\usepackage{amsfonts,amsmath,amssymb,amsthm,mathrsfs,enumerate,bm,multicol,setspace,lmodern,microtype}

\usepackage{graphicx}
\usepackage{hyperref}
\usepackage{blindtext}
\usepackage{xcolor}

\hypersetup{colorlinks=true, linkcolor=red, citecolor=blue, pdfpagemode=FullScreen}

\newtheorem{theorem}{Theorem}
\newtheorem{proposition}{Proposition}
\newtheorem{corollary}{Corollary}
\newtheorem{lemma}{Lemma}

\newtheorem{claim}{Claim}

\theoremstyle{definition}
\newtheorem{definition}{Definition}

\newtheorem{axiom}{Axiom}

\usepackage[right=1.3in, left=1.3in, bottom=1.3in, top=1.3in]{geometry}
\newcommand{\R}{\mathbb{R}}

\newcommand{\1}{\mathbf{1}}

\renewcommand{\d}{{\rm d}}

\newcommand{\calA}{\mathcal{A}}

\newcommand{\calF}{\mathcal{F}}

\newcommand{\calL}{\mathcal{L}}

\newcommand{\calU}{\mathcal{U}}
\newcommand{\calV}{\mathcal{V}}

\newcommand{\argmax}{\mathop{\rm arg~max}\limits}
\newcommand{\argmin}{\mathop{\rm arg~min}\limits}

\DeclareMathOperator{\cl}{cl}
\DeclareMathOperator{\co}{co}
\DeclareMathOperator{\cco}{\overline{{\rm co}}}

\newcommand{\df}[1]{\textit{#1}}

\usepackage{natbib}
\setcitestyle{authoryear,open={(},close={)}}

\begin{document}

\title{\textbf{Twofold Multiprior Preferences and Failures of Contingent Reasoning}\footnote{This paper subsumes the working paper ``An Axiomatic Approach to Failures in Contingent Reasoning'' by Masaki Miyashita and Yuta Nakamura, and the working paper ``Twofold Conservatism in Choice under Uncertainty'' by Federico Echenique, Luciano Pomatto, and Jamie Vinson. We are grateful to the editor and the referees for their helpful comments and suggestions.}}
\author{
Federico Echenique\thanks{California Institute of Technology; {\tt fede@caltech.edu}. Echenique thanks the National Science Foundation for financial support (Grants SES-1558757 and CNS-518941).}
\and Masaki Miyashita\thanks{Yale University; {\tt masaki.miyashita@yale.edu}}
\and Yuta Nakamura\thanks{Yokohama City University; {\tt y\_naka@yokohama-cu.ac.jp}}
\and Luciano Pomatto\thanks{California Institute of Technology; {\tt luciano@caltech.edu}}
\and Jamie Vinson\thanks{California Institute of Technology; {\tt jvinson@caltech.edu}}
}
\date{\today}
\maketitle

\begin{abstract}
We propose a model of incomplete \textit{twofold multiprior preferences}, in which an act $f$ is ranked above an act $g$ only when $f$ provides higher utility in a worst-case scenario than what $g$ provides in a best-case scenario. The model explains failures of contingent reasoning, captured through a weakening of the state-by-state monotonicity (or dominance) axiom. Our model gives rise to rich comparative statics results, as well as extension exercises, and connections to choice theory. We present an application to second-price auctions.
\end{abstract}

\newpage


\section{Introduction} \label{sec_intro}

To make coherent decisions in the face of uncertain odds is a challenge, both in policy making and in everyday life. A common response to uncertainty is to act conservatively: to choose a new option over a status quo only if there is compelling reason to do so.  In this paper, we study a formal criterion of choice under uncertainty that captures this conservative principle.
 
We study preferences over acts, where an act $f \colon \Omega \to X$ is a function assigning outcomes to states of the world. Acts may represent, for example, investments, assets, or policies. We characterize preferences $\succ$ for which the ranking $f \succ g$ admits the representation
 \[
    \min_{\mu \in C} \int u(f) \d \mu > \max_{\mu \in D} \int u(g) \d \mu.
 \]
 Here, $u$ is a utility function over outcomes, and $C$ and $D$ are sets of probabilistic ``theories,'' or  ``scenarios.'' An act $f$ is deemed superior to another act $g$ whenever $f$ provides higher expected utility than $g$, even when $f$ is evaluated under a worst-case scenario and $g$ by a best-case scenario. We call such preferences, which are in a sense doubly conservative, \textit{twofold multiprior preferences}; twofold preferences for short.
 
The most common notion of dominance between two state-contingent choices requires that one is better than the other in each state. This state-by-state principle of dominance is formalized as a monotonicity axiom, and often regarded as a basic tenet of rationality; but there is mounting empirical evidence showing that people may fail to recognize or choose a dominant act. The evidence spans several different economic environments, including  voting \citep{esponda2014hypothetical,esponda2019contingent}, auctions \citep{kagel1987information,kagel1993independent,charness2009origin,li} and matching problems \citep{chen2006school,echenique2016clearinghouses,rees2017mistaken,chen2019self}. There are also recent experimental findings that cast doubts on monotonicity as a consideration in choice under uncertainty \citep{schneider2019experimental}.

Twofold preferences assume an agent who understands the possible consequences of an act, but may not be able to fully reason in terms of the underlying state space.  Without the ability to perform contingent reasoning, different scenarios may be applied to evaluate different choices, which can explain why agents violate monotonicity, and fail to choose a dominant action. For instance, in a second-price auction, overbidding would be ``rationalized'' compared to truth-telling if agents make an \emph{as if} assumption that the more aggressively she bids, her opponents are persuaded to bid lower values (see Section~\ref{sec_auctions}).

 Twofold preferences  have been proposed in philosophy by \cite{kyburg1983rational}, as a  theory of choice that is more robust than subjective expected utility; and \cite{nozick1969newcomb} proposed strategy-dependent reasoning as a contrast to contingent reasoning to explain Newcomb's paradox. In decision theory, the theoretical structure of interval orders introduced by \cite{fishburn1973interval} also uses twofold preferences. More recently, the logic behind twofold preferences has found fruitful applications in mechanism design. \cite{li} introduced the notion of \df{obvious strategy-proofness}, a solution concept which requires truthtelling to be optimal  even when its consequences are evaluated according to the worst-case scenario, but the consequences of a false report are evaluated according to the best-case scenario. Obvious strategy-proofness is the subject of a recent, growing, literature in mechanism design. 

 Indeed, a key idea in Li's obvious strategy proofness is that agents have a limited ability towards contingent reasoning.\footnote{It is worth emphasizing that \cite{esponda2019contingent} designed laboratory experiments to detect the cause of anomalies and deduced that most anomalies can be explained by failures in contingent reasoning.} This limitation is one of the foundations provided by Li for obvious strategy proofness in mechanism design.\footnote{Twofold preferences capture obvious dominance when $C$ and $D$  comprise all beliefs over the state space, but is a more flexible model. For instance, in a second-price auction, all bidding strategies are undominated with respect to Li's rule, but twofold can yield a reasonable  non-trivial range of undominated bidding strategies.} In our setting, limited contingent reasoning means a limitation to reason state-by-state. Our axiomatization captures this through a weakened monotonicity axiom. The standard notion of monotonicity states that if $f(\omega) \succ g(\omega)$ for all $\omega$, i.e.\ if the outcome given by $f$ is better than the outcome given by $g$ in every state, then the ranking $f \succ g$ must follow. We show that twofold preferences satisfy a weakening of this condition. In fact, a twofold preference satisfies the stronger state-by-state monotonicity property if and only if it collapses to subjective expected utility. As we discuss in Section~\ref{sec:discussion}, twofold preferences model agents  who cannot formulate a theory of how acts result in different outcomes depending on the realized state; they have a \emph{practical} understanding of each act, not a theoretical one. 

Our axiomatization requires constant acts to be easier to compare to other acts. We maintain completeness over constant acts \citep{bewley2002knightian}, as well as independence when mixing with constant acts \citep[C-independence, see][]{gilboa1989maxmin}. And our main axiom  says that, whenever the decision maker is unwilling to rank neither acts $f$ nor $g$ against a constant act $x$, then the two acts must be incomparable. This implies the existence of a certain frame of reference for any twofold conservative preference: If the decision maker can compare $f$ and $g$, she must be able to justify her choice by pointing to a constant act that she can compare $f$ or $g$ to. 

Finally, a novel aspect of  twofold preference is that it combines aversion to the ambiguity in the alternative act $f$ with a potential preference for the ambiguity in the status quo. As far as we know, our paper is the first to point out that attitudes towards ambiguity could have a differential impact on a status quo and its competing options. We exploit this structure to develop novel comparative statics results in Section~\ref{sec_cs}, and explore an alternative model of twofold preferences (a maxmin model) in which the preference for ambiguity in the status quo is dropped (Section~\ref{sec_maxmin}).

The closest model to twofold preferences, among the models of conservatism in choice under uncertainty, is due to \cite{bewley2002knightian}. A key difference with Bewley's model is that his agents can compare two acts state-by-state and prior-by-prior, two instances of contingent reasoning that are not assumed in twofold preferences. Compared to Bewley's model, the criterion described by twofold preferences may seem  simplistic: making the decision metric so obvious that very few decisions are actually made. However, we show that twofold preferences have many of the same useful properties of Bewley preferences. In particular, under some conditions, any choice function that is weakly rationalizable by a Bewley preference can also be weakly rationalizable by a twofold preference, yet the converse is false (Section~\ref{sec:choice}). Thus, arguably, twofold conservative preferences can explain a strictly larger set of decision makers. In addition, the result by \cite*{gilboa2010objective}, which shows how Bewley preference can be extended into a complete maxmin preference, can equivalently be formulated by starting from a twofold preference. Details are in Section~\ref{sec_extension}.


\subsection{Related Literature}

 Our representation is a special case of an \emph{interval order}, a type of preference relation first studied by \cite{fishburn1970intransitive,fishburn1973interval}.\footnote{\cite{bridges2013representations} is an introduction to the mathematics of interval orders.} This is easily seen by considering, for each act $f$, the functionals $\underline{U}(f) = \min_{\mu \in C} \int u(f) \d \mu$ and $\overline{U}(f) = \max_{\mu \in D} \int u(f) \d \mu$. Then, as needed for an interval order, each act $f$ is associated with an interval $[\underline{U}(f),\overline{U}(f)]$, and $f$ is preferred to $g$ if and only if it holds that $\underline{U}(f) > \overline{U}(g)$.
 
 Within the literature on interval orders, \cite{nakamura1993subjective} is the contribution closest to our paper. He proposes an interval order representation inspired by \citeauthor{schmeidler1989subjective}'s (\citeyear{schmeidler1989subjective}) theory of Choquet expected utility, where $\underline{U}$ and $\overline{U}$ are Choquet integrals with respect to two capacities. Not only the representations but also the type of axioms involved are quite different: While one of his main axioms is a complex multisymmetry assumption, our axioms are instead specifically tailored to an Anscombe-Aumann framework and rely on familiar concepts, such as monotonicity, the idea of hedging against uncertainty, and a preference for sure over uncertain outcomes. 

 A key motivation behind our paper is the notion of obvious dominance in mechanism design \citep{li}. This has been studied axiomatically before, by \cite{zhang2017bounded}. They model obvious dominance through a complete preference relation, and obtain a weighted min-max criterion. Our point of departure is different:\ we think of obvious dominance as necessitating an incomplete criterion. In our model a status quo is only abandoned in favor of an alternative if there is sufficient reason to do so, and therefore a mechanism is robustly implementable if a deviation would be adopted in favor of the proposed equilibrium, even if the deviation were the status quo. 

 The recent work of \cite{valenzuela2020subjective} also uses preference incompleteness to model complexity in decision problems. His focus is on a measure of complexity that depends on the cardinality of the partition generated by each act. The representation is distinct from ours, but shares the spirit of accounting for preference incompleteness through the multiple evaluations of each act.
 
 A number of recent papers relax monotonicity. \cite{ellis2017correlation} study a weakening of monotonicity in a model of correlation misperception. They enrich the standard Anscombe-Aumann to allow for the possibility that a decision maker may distinguish between the left and right sides of the expression $h = \alpha f + (1-\alpha)g$, even though the two refer to same act. \cite{saponara2020revealed} models a decision maker who has limited understanding of the state space, and evaluates each act with respect to a class of partitions of events. \cite{puri2020} presents a model of choice over lotteries that incorporates the cognitive costs of contemplating lotteries with large support. Her model predicts violation of first order stochastic dominance, which is analogous to violations of monotonicity in this paper. 
 
 Several well-known models of choice under uncertainty express a degree of caution or conservatism, for example maxmin preferences \citep{gilboa1989maxmin}, and variational preferences \citep*{maccheroni2006ambiguity}. Twofold preferences differ from these in providing an incomplete ranking of acts. In Section~\ref{sec_extension} we offer a connection through an extension result from twofold preferences to complete preferences, in the spirit of \cite*{gilboa2010objective}, \cite*{cerreia2016objective},  \cite*{cerreia2020rational}, and \cite*{frick2021objective}.
 
 A number of authors have proposed models of incomplete preferences under uncertainty that can be seen as alternatives to Bewley's theory, including \cite{nascimento2011class}, \cite{minardi2015preferences}, \cite{hill2016incomplete}, and \cite{faro2015variational}. Incomplete preferences have also been studied experimentally, by  \cite{danan2006preferences} and \cite{cettolin2019revealed}, among others.


\section{Model and Representation} \label{sec_preliminary}

 We adopt the framework of \cite{anscombe1963definition}, the workhorse of modern work on decisions under uncertainty. Given are a finite set $\Omega$ of \emph{states of the world} and a set $X$ of \emph{outcomes}. The set $X$ is a convex subset of a vector space.  An \emph{act} is a function $f \colon \Omega \to X$. We denote by $\calF$ the set of all acts, and identify an outcome $x \in X$ with the constant act whose outcome is $x$ in every state. Given $f,g \in \calF$ and $\alpha \in (0,1)$, the mixture $\alpha f + (1-\alpha) g$ is the act defined as $\alpha f(\omega) + (1-\alpha) g(\omega)$ in each state. Finally, $\Delta(\Omega)$ represents the space of probability measures over $\Omega$.

 A binary relation $\succ\ \subseteq{\calF}\times{\calF}$ is a set of ordered pairs of acts describing a partial ranking over acts. We write $f\succ g$  when $(f,g)\in {\succ}$, as is standard. When $f \not \succ g$ and $g \not \succ f$ we say that $f$ and $g$ are not comparable, and write $f\bowtie g$.
 
The relation $\succ$ can be interpreted as the preference comparisons that the decision maker is willing to make with sufficient conviction. The relation $\succ$ can also be interpreted as describing choices under a status quo \citep{bewley2002knightian}. In that case, $f\succ g$ means that $f$ is chosen from the set $\{f,g\}$ when $g$ is the default, or status quo, option; and $f\bowtie g$ means that $f$ is not chosen, and the status quo not abandoned. See Section~\ref{sec_rational} for an explicit model of choice under a status quo.

Throughout the paper, we will focus on binary relations $\succ$ that, when restricted to constant acts, are the asymmetric part of a complete and transitive binary relation $\succsim$. We will then denote by $\sim$ the symmetric part of $\succsim$ over constant acts. Thus, $\sim$ will only be used to compare constant acts.


\subsection{Representation}\label{sec:modelandpreviewofmainresults}

 A decision maker, named Alice, must make a choice between two acts. Under subjective expected utility, the standard model of choice under uncertainty, Alice is endowed with a utility function $u \colon X \rightarrow \mathbb{R}$, and a \emph{prior} $\mu\in\Delta(\Omega)$, so that she ranks act $f$ over $g$ if and only if 
\begin{equation}\label{eq:SEU}
 \int u(f) \d \mu > \int u(g) \d \mu .
\end{equation} 
 More precisely, we call a binary relation $\succ$ a \emph{subjective expected utility preference} if there is a pair $(u,\mu)$, with $u \colon X\to \R$ non-constant and affine and $\mu\in\Delta(\Omega)$ that represents $\succ$.

 A more conservative choice criterion was studied by \cite{bewley2002knightian}. In his theory, in a choice between two acts $f$ and $g$, the act $g$ is special: it is a default, or \textit{status-quo} option. Alice acts conservatively, meaning that she will only choose $f$ over the status quo $g$ if there are good reasons to do so. In her decision, Alice evaluates each option by means of a set $C \subseteq \Delta(\Omega)$, with each $\mu \in C$ representing a different theory about the states of the world. An act $f$ is chosen over the status  quo act $g$ if and only if
 \begin{equation}\label{eq:bewley}
    \int u(f) \d \mu > \int u(g) \d \mu \text{~~for all~} \mu \in C.
 \end{equation}
 More formally, a binary relation $\succ$ over acts is termed a \emph{Bewley preference} if there exists an (affine and non-constant) utility $u \colon X \to \R$ and a closed convex set $C$ such that~\eqref{eq:bewley} holds if and only if $f \succ g$. 

 In this paper we introduce a new decision criterion:
\begin{definition} \label{def_rep}
 A binary relation is a \emph{twofold multiprior preference} (or, a \emph{twofold preference} for simplicity) if there exist a non-constant affine function $u\colon X \to \R$, and two closed convex subsets $C,D \subseteq \Delta (\Omega)$ with non-empty intersection such that for all $f,g \in \calF$,
 \begin{align} \label{eq_rep}
 f \succ g \Longleftrightarrow \min_{\mu \in C} \int u(f) \d \mu > \max_{\mu \in D} \int u(g) \d \mu.
\end{align}
The relation $\succ$ is in addition \emph{concordant} if $C = D$.
\end{definition}

 Consider first the case where $\succ$ is concordant. An act $f$ is deemed preferable to another act $g$ when $f$ gives an higher expected payoff, according to the worst-case scenario $\mu \in C$, than $g$ gives in the best-case scenario. The set  of scenarios is subjective and revealed from the decision maker's choices. In this representation, each act $f$ is evaluated in terms of the range $\{\int u(f) \d \mu : \mu \in C\}$ of expected payoffs it can lead to, and an act $f$ is deemed better than another act $g$ only when the interval induced by $f$ is disjoint from, and dominates, the interval induced by $g$.\footnote{In the representation, closedness and convexity of the sets $C$ and $D$ are assumptions that guarantee stronger uniqueness properties for the representation. The essential assumption is that $C$ and $D$ have nonempty intersection, which ensures that $\succ$ is an asymmetric binary relation.}
 
 
 
When the preference $\succ$ is not concordant, different sets of theories are applied to the two acts $f$ and $g$.
This can be an appropriate modeling choice when, for example, $g$ is a status-quo medical treatment and $f$ a new, alternative, treatment.
In such situations, a more conservative decision maker may want to apply a more demanding standard to the new treatment, all other things being equal.
This can be formally captured by enlarging the set $C$ while keeping the size of $D$ fixed.

\begin{figure}[t!]
\centering
\includegraphics[width=15cm]{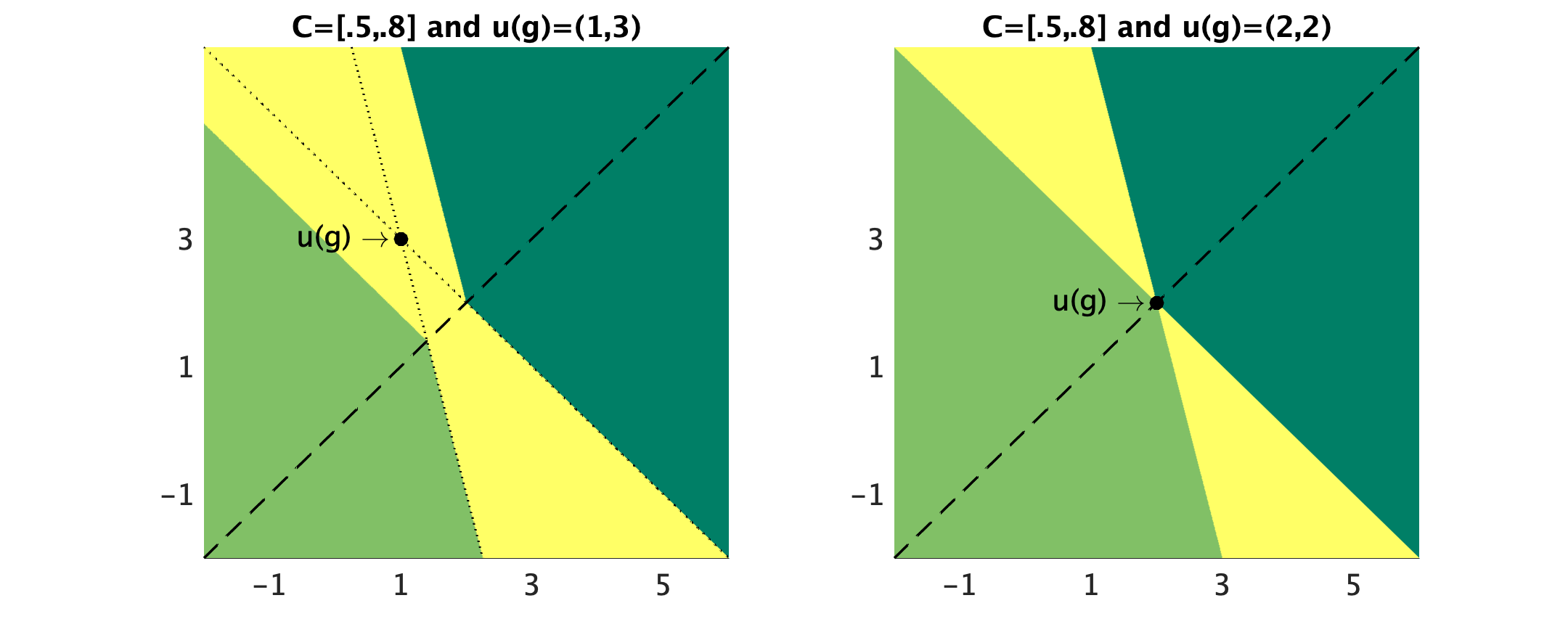}
\caption{Upper contour sets (dark green), lower contour sets (light green), and incomparable sets (yellow) are displayed for a twofold multiprior preference with the concordant belief set $C=[.5,.8]$.
The status-quo point is $(1,3)$ in the left, and it is $(2,2)$ in the right.
All boundaries are included in the yellow regions by the openness of $\succ$.}
\label{fig_contour1}
\end{figure}

\begin{figure}[t!]
\centering
\includegraphics[width=15cm]{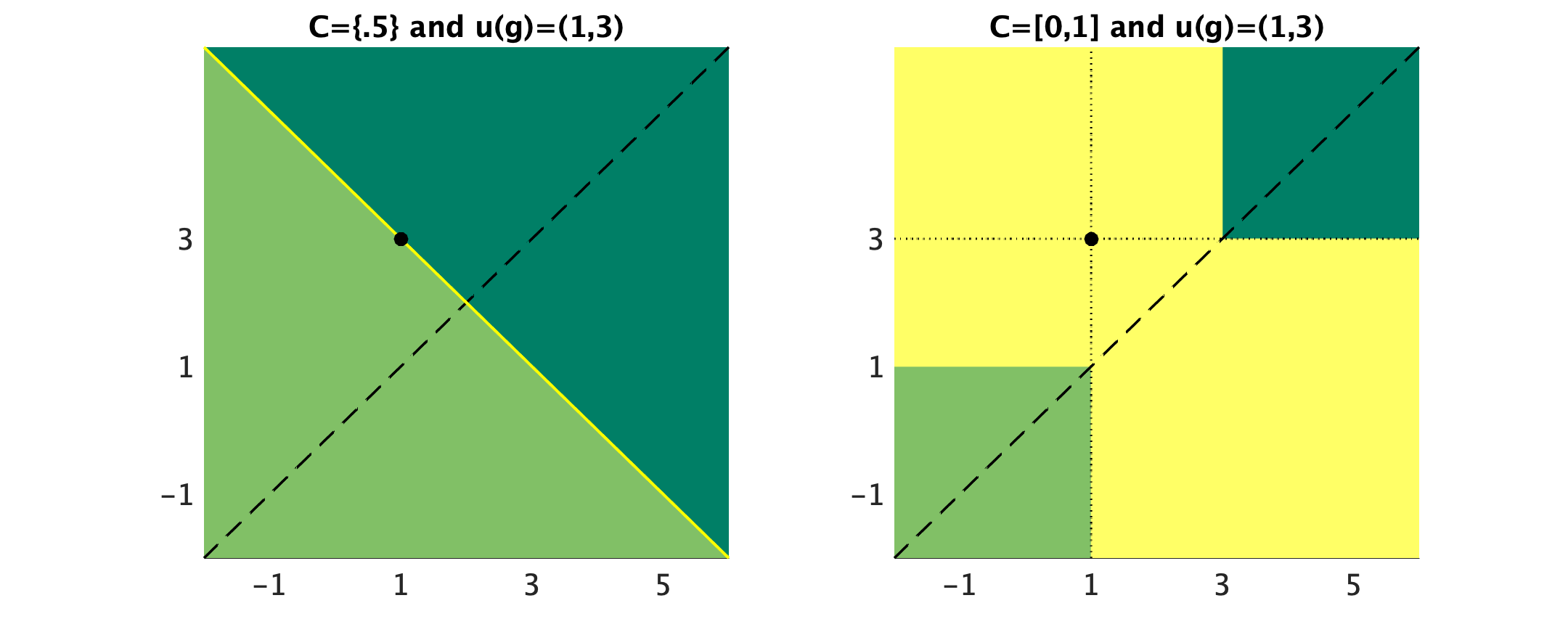}
\caption{The same as the previous figure, except the belief sets to be considered.}
\label{fig_contour2}
\end{figure}

Contour sets are useful to grasp the nature of our representation.
In Figures \ref{fig_contour1} and \ref{fig_contour2}, we show the upper and lower contour sets of a twofold multiprior preference in some simple examples.
In these examples, there are two states, $\omega_1$ and $\omega_2$, and each axis represents the utility $u(f(\omega))$ given by the act $f$ in the corresponding state.
The status-quo act $g$ is displayed as a black bullet, whose upper and lower contour sets are displayed as dark green and right green regions, respectively.
The remaining yellow regions correspond to the set of utility acts that are incomparable to $g$.
Note that upper and lower contour sets are open convex sets. 
Since we focus on the concordant case $C = D$, they are also located symmetrically around $g$.

Figure \ref{fig_contour1} considers the belief set $C=[.5,.8]$, while two panels differ in the locations of $g$.
When the status quo is not on the $45^\circ$ line (see the left panel), it is contained in a ``thick'' incomparable region.
In contrast, when the status quo is on the $45^\circ$ line (see the right panel), the region of incomparability shrinks towards the status-quo point, so that acts that are close to the reference point are comparable to $g$; indeed here the contour sets coincide with those generated by a Bewley preference with the same belief set.
Importantly, the thickness of the incomparable region depends on how asymmetric the reference point is, i.e., the incomparable region expands as the reference point moves away from the $45^\circ$ line.

Figure \ref{fig_contour2} illustrates the effect of different belief sets, subjective expected utility corresponds to the special case when $C$ is a singleton (see the left panel).
Obvious dominance emerges when $C$ consists of all possible beliefs (see the right panel).
Evidently, the larger the set $C$ becomes, the larger the region of incomparability.

\subsection{Discussion}\label{sec:discussion}

 We now present a literal, or ``psychological,'' interpretation of twofold multiprior preferences. Our purpose is to explain how twofold preferences depart from Bewley's model, their direct intellectual predecessor. The rest of the paper examines the behavioral content of this class of preferences, and how it compares to other models of choice under uncertainty.
 
Twofold preferences describe a decision maker who assigns to each act a set of possible payoff consequences---consequences which we term \textit{evaluations}. Alice, our decision maker, does not fully think in terms of states of the world, and has only a limited understanding of how acts result in different outcomes depending on the realized state. She has a practical understanding of each act; not a theoretical one. Alice knows that an act is associated with a range of possible payoff consequences, but in making comparison between acts, she finds it difficult to trace how the consequence of her action vary across states.
 
 Examples of such situations are common. The ancient Romans added wine to water for sanitary reasons, even though they had no knowledge of micro-organisms, or of how pathogens are affected by alcohol.\footnote{Alice's practical knowledge illustrates a very common situation. The subject of wine in Rome alone provides many examples: Cato The Elder recommended careful cleaning of wine-making instruments, and the physician Galen used wine to clean gladiators' wounds. The ancient Romans had no theoretical understanding of fermentation or infections.} In modern days, a physician may know how a certain drug interacts with a number of common health conditions, without necessarily having complete knowledge of the underlying biological process. As another example, consider an agent who is participating in an allocation mechanism such as deferred acceptance. Even though she may have a general understanding of the rules of  the mechanism and its main properties, she may not be able to pin down the exact allocation that is obtained when a certain profile of preferences is reported.\footnote{Now it may seem contradictory to raise these issues in a model that features an explicit (objective) state space: there is an asymmetry in how the analyst (us) identifies acts with state-contingent choices, while Alice has a simpler representation of the objects she can choose among.\label{page:forR2}}

In our discussion, we refer to a probability distribution $\mu\in\Delta(\Omega)$ as a \textit{theory}, and to the expected utility $\int u(f) \d \mu$ as an \textit{evaluation}. Bewley's  decision maker does not know the theory that governs the resolution of uncertainty: an ignorance which is reflected in a set $C\subseteq \Delta(\Omega)$ of possible theories. Bewley's decision maker has access to $C$, and will choose an act $f$ over $g$ if the evaluation according to any theory in $C$ ranks $f$ over $g$. Observe that this is a key aspect of the representation: the ability to evaluate two acts for each possible theory. It follows then that Bewley's Alice understands how acts map states of the world into consequences. Knightian uncertainty for her is about not knowing the right theory, not about a poor understanding of how states map into consequences. 

Our Alice, in contrast, cannot use theories to compare two acts, but she still has access to the set of evaluations for each act. Alice associates a set $\{\int u(f)\d \mu:\mu\in C \}$ of payoff consequences to $f$. She suffers from the same status-quo bias as Bewley's Alice, so she is only willing to give up $g$ in favor of $f$ if $\min \{\int u(f) \d \mu:\mu \in C \}> \max \{\int u(g)\d \mu:\mu\in C \}$. But since she does not have access to the set of theories $C$, she finds it difficult to compare $f$ and $g$ for each individual element of $C$. Instead, she compares the two acts based on the range of possible evaluations that an act may result in. Her conservatism then results in our representation.\footnote{In this discussion we focus on concordant preferences.}

To sum up, we think of twofold preferences as a natural model of conservatism. In the spirit of Bewley, but pushing his ideas one step further. Twofold preferences arise when the decision maker's ignorance is expressed in terms of uncertain payoff consequences instead of uncertain theories.  

\subsection{An Illustration in Second-price Auctions}\label{sec:illustrativeexample}

The difference between conservatism in Bewley's model and in twofold preference can yield starkly different predictions about a decision maker's choice. As an illustration, we discuss the second-price auction. In particular, we compare Bewley's model with the notion of obvious dominance, introduced by \cite{li}, and used by him to account for empirically observed deviations from dominant strategies in second-price auctions. We then show how twofold preference captures similar behavior to obvious dominance in second-price auctions, but with added flexibility. A formal model of second-price auctions is presented and studied in Section~\ref{sec_auctions}.

In the present decision-theoretic framework, Li's decision criterion corresponds to the special case of concordant twofold representation with the universal belief sets  $C=D=\Delta(\Omega)$, i.e.,
\begin{align} \label{eq_od}
f \succ g &\Longleftrightarrow \min_{\mu \in \Delta(\Omega)} \int u(f) \d \mu > \max_{\mu \in \Delta(\Omega)} \int u(g) \d \mu.
\end{align}
In other words, $f$ {\it obviously dominates} $g$ if and only if the best payoff consequence that can result from $f$ is higher than the worst payoff consequence from $g$, as described by the contour sets in the right panel of Figure~\ref{fig_contour2}.

Bewley’s decision maker faces uncertainty over not knowing the right theory, while she has a theoretical understanding of how each act will be mapped to its payoff consequences.
In the context of a second-price auction, she does not know which $\mu \in \Delta(\Sigma)$ describes the distribution of opponents’ bids (states are bids); but she can evaluate the consequences of any given strategy contingent on each $\mu$. Since truth-telling is a dominant strategy in a second-price auction, Bewley's decision maker can thus identify truth-telling as an optimal choice, even without the right theory of the bids chosen by her opponents.

In contrast, obvious dominance postulates a decision maker who only has a practical understanding of each bidding strategy. She has access to the possible range of evaluations $\{\int u(f) \d \mu: \mu \in \Delta (\Omega)\}$ for each strategy $f$, but does not know how exactly each theory $\mu$ is associated with the payoff consequence of $f$.
Hence, she may apply different theories in comparing different strategies. For example, she may think {\it as if} her opponents would bid lower when she bids aggressively; and, as a result, the payoff ranges of different strategies may overlap. On the basis of (\ref{eq_od}), this means that there is no effective way to eliminate any strategy from consideration.

Twofold preferences capture a decision criterion that is qualitatively similar to obvious dominance. Our decision maker can only associate the range of possible evaluations $\{\int u(f)\d \mu:\mu\in C \}$ to each $f$. Truth-telling is not better than a deviating strategy, unless their payoff ranges are disjoint. But twofold preferences can be more flexible in the set of possible theories $C$ (or $D$), and thus may provide quantitatively sharper predictions about which deviations are likely to occur. As such, in Section~\ref{sec_auctions}, we see that the set of strategies that are undominated in terms of the twofold representation is decreasing in belief sets, while obvious dominance does not preclude any behavior.

\section{Representation Theorem} \label{sec_main}

 We now proceed with an axiomatic characterization of twofold multiprior preferences. Our first three axioms are standard.

\begin{axiom} \label{axiom_str}
 $\succ$ is asymmetric and transitive. The restriction of $\succ$ to $X$ is non-trivial and negatively transitive.
\end{axiom}

\begin{axiom} \label{axiom_cts}
 For all $f,g,h \in \calF$, the sets $\{\alpha \in [0,1]: \alpha f + (1-\alpha) g \succ h\}$ and $\{\alpha \in [0,1]: h \succ \alpha f + (1-\alpha) g\}$ are open.
\end{axiom}

\begin{axiom} \label{axiom_cindp}
 For all $f,g \in \calF$, $x \in X$, and $\alpha \in [0,1]$, $f \succ g$ if and only if $\alpha f + (1-\alpha) x \succ \alpha g + (1-\alpha) x$.
\end{axiom}

 Axiom~\ref{axiom_str} collects standard assumptions on preferences. It implies, in particular, that the restriction of $\succ$ to $X$ is the strict part of a binary relation $\succsim$ that is complete and transitive. Axiom~\ref{axiom_cts} is a standard Archimedean continuity condition. Axiom~\ref{axiom_cindp}, termed \textit{C-independence}, is a well-known weakening of Anscombe and Aumann's independence axiom due to \cite{gilboa1987expected} and \cite{gilboa1989maxmin}.

The next axiom concerns the convexity of upper and lower contour sets.

\begin{axiom} \label{axiom_conv}
For all $x \in X$, the sets $\{f \in \calF: f \succ x\}$ and $\{f \in \calF : x \succ f\}$ are convex.
\end{axiom}

Following \cite{schmeidler1989subjective}, convexity of the upper contour sets $\{f \in \calF : f \succ x\}$ expresses aversion to uncertainty. It means that hedging, interpreted as the convex combination of acts, never makes the decision maker change their mind in abandoning a constant act $x$ for the alternative act $f$.

Convexity of lower contour sets is a less common assumption, and expresses a preference for uncertainty when uncertainty concerns the status quo.\footnote{
We present an alternative model of twofold preferences, in which we depart from Axiom~\ref{axiom_conv}, in Section~\ref{sec_maxmin}.}
If the decision maker is willing to abandon either act $f$, or $g$, in favor of an alternative constant act $x$, then hedging between $f$ and $g$ does not provide for a more appealing status quo, and would not overturn her decision. Moreover, under a twofold multiprior preference, hedging between two status-quo acts $f$ and $g$ can lead the decision maker to abandon the status quo even if she was not willing to do so for either of the two original acts. That is, a twofold multiprior preference can exhibit the rankings $x \nsucc f$, $x \nsucc g$ and at the same time $x \succ \alpha f + (1-\alpha)g$. Thus, hedging can make the decision maker strictly more willing to abandon the status quo.\footnote{Equivalently, for a twofold multiprior preference the sets $\{f \in \calF : x \nsucc f\}$ may not be convex. This property does not follow from Axiom~\ref{axiom_conv} alone, but it can be easily verified from Definition \ref{def_rep}.}

 The next axiom states that, whenever the decision maker is unwilling to rank neither $f$ nor $g$ against a constant act $x$, then the two acts must be incomparable. 

\begin{axiom} \label{axiom_contagion}
For all $f,g \in \calF$ and $x \in X$, if $f \bowtie x$ and $g \bowtie x$, then $f \bowtie g$.
\end{axiom}

Axiom~\ref{axiom_contagion} captures the idea that, for the decision maker, a choice between an act and a constant act is easier than a choice between two arbitrary acts. In other words, an agent who can compare two acts must also be able to compare them to a constant act. 

Next, we turn to the standard postulate of \emph{monotonicity}, which relies on state-wise dominance: $f(\omega) \succ g(\omega)$ for all $\omega$, implies the ranking $f \succ g$. We are interested in preferences that may violate state-wise dominance, as we have in mind an agent who may be unable to fully reason state-by-state. Thus our next two axioms are a weakening of monotonicity.

\begin{axiom} \label{axiom_equiv}
For all $f,g \in \calF$ and $x \in X$, if $f(\omega) \sim g(\omega)$ for all $\omega$, then $f \succ x$ implies $g \succ x$, and $x \succ f$ implies $x \succ g$.
\end{axiom}

\begin{axiom} \label{axiom_mono}
For all $f,g \in \calF$ and $x,y \in X$, if $x \succ f(\omega) \succ y$ for all $\omega$, then $x \succ f \succ y$.
\end{axiom}

 In weakening the usual state-by-state monotonicity axiom, Axioms \ref{axiom_equiv} and \ref{axiom_mono} relate to the motivation behind our representation. We have in mind an agent who finds complex state-wise comparisons cognitively challenging, and thus has only a partial understanding of the relation between states and outcomes. Axiom \ref{axiom_equiv} reflects this intuition: In order to verify the hypothesis that $x \succ f(\omega) \succ x$ holds for all $\omega$, the decision maker does not need to fully understand how the act $f$ vary across states, only the set of outcomes $\{ f(\omega) : \omega \in \Omega\}$ it can lead to.\label{page:mon}
 
 Axiom~\ref{axiom_mono} allows us to separate the decision maker's tastes over outcomes from other aspect of her preference. It is automatically satisfied in a number of cases, as when $X = \mathbb{R}$ and the preference is strictly increasing over $X$, or when $X$ is the set of lotteries over two outcomes. It may be more difficult to satisfy when the set of outcomes is very rich (for example, lotteries over high dimensional commodities), in which case verifying $f(\omega) \sim g(\omega)$ in every state may involve complicated state-by-state comparisons.

 Our first theorem is the main result of the paper. It provides an axiomatic foundation for twofold preferences, as well as a uniqueness property of the representation. 

\begin{theorem} \label{thm_main}
A binary relation $\succ$ satisfies Axioms \ref{axiom_str}--\ref{axiom_mono} if and only if it admits a twofold multiprior representation.
Moreover, $u$ is unique up to positive affine transformations, and $C$ and $D$ are unique.
\end{theorem}

 The proof of Theorem~\ref{thm_main} is in Appendix~\ref{appendix_main}. Here, we provide a sketch, and give some intuition on the relation between the axioms and the representation. By Axioms \ref{axiom_str}--\ref{axiom_cindp}, the restriction of $\succ$ to $X$ satisfies the conditions of the von-Neumann Morgenstern theorem, and thus it is represented by an affine function $u\colon X \to \R$. To extend the representation to general acts, we associate to each $f \in \calF$ 
 the {\it minimal} and {\it maximal} utilities that are assigned to constant acts incomparable with $f$. Formally, we set
\begin{align} \label{eq_def_util}
\underline{U}(f) = \inf_{x \in X} \{u(x): f \bowtie x\} \ \text{ and } \ 
\overline{U}(f) = \sup_{x \in X} \{u(x): f \bowtie x\}.
\end{align}

 The essential idea is that $\underline{U}$ and $\overline{U}$ represent $\succ$ in the sense that $f \succ g$ if and only if $\underline{U}(f) > \overline{U}(g)$. It is an implication of transitivity that $f \succ g$ whenever $\underline{U}(f) > \overline{U}(g)$. In proving the converse direction, a key step is to establish that for every pair of acts, there exist constant acts $x$ and $y$ such that $f \succ x \succ y \succ g$ (see Lemma~\ref{lem_calibration} in the appendix). This property is a defining feature of our representation, pertaining to the specific way our decision maker compares uncertain to constant acts. It is in general not satisfied by other incomplete relations, such as Bewley's preferences.

\begin{figure}[t!]
\begin{center}
\includegraphics[width=16cm]{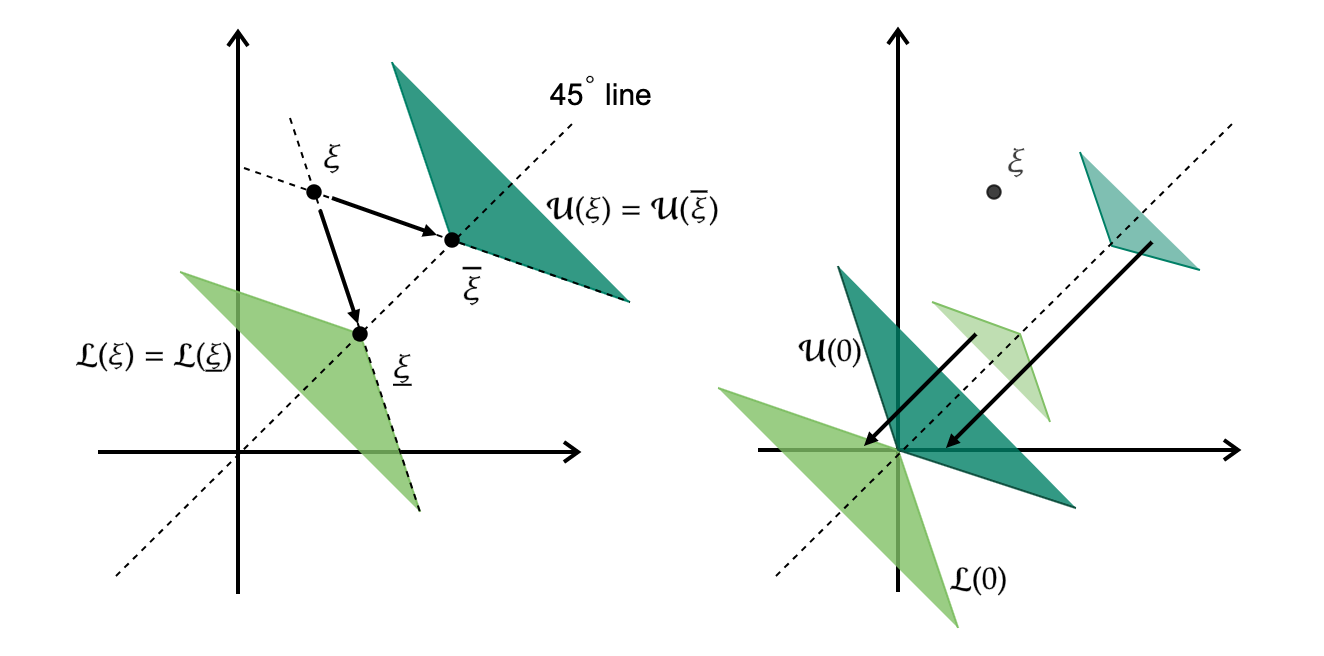}
\caption{The contour sets of a utility act $\xi$ are shifted to the origin in two steps.}
\label{fig_sketch1}
\end{center}
\end{figure}

 The second step of the proof consists in deriving the integral representations for $\underline{U}$ and $\overline{U}$. As standard, we associate each act $f$ to the utility-valued act $\xi \equiv u(f)$, and define the functionals $\underline{T}(u(f)) = \underline{U}(f)$ and $\overline{T}(u(f)) = \overline{U}(f)$ on $\R^\Omega$. It is an implication of Axiom~\ref{axiom_equiv} that $\underline{T}$ and $\overline{T}$ are well defined. Using Axioms \ref{axiom_cindp}--\ref{axiom_mono} we then establish a number of properties of $\underline{T}$ and $\overline{T}$, including restricted forms of additivity and monotonicity. The \cite{gilboa1989maxmin} representation theorem is not applicable in the absence of the full monotonicity, so we develop alternative arguments that recover the sets $C$ and $D$ by studying the upper and lower contour sets, as in \cite{bewley2002knightian}. Given $\xi \in \mathbb{R}^\Omega$, we consider the contour sets
\begin{align*}
\calU(\xi) = \left\{\zeta \in \R^\Omega: \underline{T}(\zeta) \ge \overline{T}(\xi)\right\} \ \text{ and } \
\calL(\xi) = \left\{\zeta \in \R^\Omega: \underline{T}(\xi) \ge \overline{T}(\zeta)\right\}.
\end{align*}

From the properties of $\overline{T}$ and $\underline{T}$, one can show that $\calU(\xi)$ and $\calL(\xi)$ are closed and convex cones. Moreover, they can be shifted to the origin through two steps that are illustrated in Figure~\ref{fig_sketch1}. First, we show that they can be identified with the contour sets of some elements on the diagonal, i.e.\ we can identify constant vectors $\overline{\xi}$ and $\underline{\xi}$ on the $45^\circ$ line such that $\calU(\xi) = \calU(\overline{\xi})$ and $\calL(\xi) = \calL(\underline{\xi})$ hold.
 We next prove that the contour sets of $\overline{\xi}$ and $\underline{\xi}$ are shifted linearly toward the origin, in the sense that $\calU(\overline{\xi}) = \calU(0) + \overline{\xi}$ and $\calL(\underline{\xi}) = \calL(0) + \underline{\xi}$.
 These arguments imply that it is enough to characterize $\calU(0)$ and $\calL(0)$ in order to recover contour sets of general acts. In the final step, the sets $C$ and $D$ are recovered through duality arguments.

\section{Comparative Statics} \label{sec_cs}

\subsection{Concordance and Attitudes Towards Uncertainty}\label{sec:concordance}

 The next question in our study of the twofold representation concerns the nature of the sets $C$ and $D$ of priors. The literal interpretation of the representation is that, in deciding whether to abandon a status quo, the set of priors that generate the payoff values of the status quo may in principle differ from the set behind the values of the new alternative to be adopted. It turns out that the two sets of priors capture the decision maker's attitude towards uncertainty.

The idea is simple. In deciding whether to choose an act $f$ over a constant act $x$, Alice has to compare the worst-case evaluation $\min_{\mu \in C}\int u(f) \d \mu $ with $u(x)$. So $C$ captures Alice's aversion to uncertainty. In contrast, in deciding whether to choose a constant act $x$ over an uncertain act $f$, Alice has to compare $u(x)$ with $\max_{\mu \in D} \int u(f)\d \mu$, so $D$ captures her inclination to keep the uncertain act instead of switching to a constant one. In consequence, the set $D$ captures the extent to which Alice has stopped worrying, and learned to love ambiguity.

To formalize this idea,  we study pairs of acts that perfectly hedge against each other. Following \cite{siniscalchi2009vector}, we say that two acts $f$ and $g$ are \emph{complementary} if
\begin{align*}
\frac{1}{2}f(\omega) + \frac{1}{2}g(\omega) \sim \frac{1}{2}f(\omega') + \frac{1}{2}g(\omega') \text{~for all~} \omega, \omega' \in \Omega.
\end{align*}

When hedging between two complementary acts $f$ and $g$, the resulting mixture $\frac{1}{2}f + \frac{1}{2}g$ is in every state equivalent to a constant act, so that any variation in utility between the two acts cancels out. Complementary acts move in ``opposite'' directions, as function of the state---one act performs better in some states but worse in others---and this trade-off can make a choice between the two difficult. Hence, if the decision make manages to rank $f$ and $g$, a choice that involves their mixture might become an easier task, as the mixture is equivalent to a constant act. Our next axioms are based on this idea:

\begin{axiom}\label{sym1}
 If $f$ and $g$ are complementary, then $f \succ \frac{1}{2}f + \frac{1}{2}g$ implies $\frac{1}{2}f + \frac{1}{2}g \succ g$.
\end{axiom}

\begin{axiom}\label{sym2}
If $f$ and $g$ are complementary, then $\frac{1}{2}f + \frac{1}{2}g \succ g$ implies $f \succ \frac{1}{2}f + \frac{1}{2}g$.
\end{axiom}

Axioms~\ref{sym1} and~\ref{sym2} differ only in which uncertain prospects, $f$ or $g$, is assumed comparable to the mixture $\frac{1}{2}f + \frac{1}{2}g$. Axiom~\ref{sym1} expresses a form of dislike for uncertainty. The premise $f \succ \frac{1}{2}f + \frac{1}{2}g$ means that, despite the uncertainty inherent in $f$, the decision maker is willing to choose $f$ over $\frac{1}{2}f + \frac{1}{2}g$. Then, Axiom~\ref{sym1} says, the conclusion that $\frac{1}{2}f + \frac{1}{2}g \succ g$ follows. For an uncertainty-averse agent, the latter choice comes easier than the former. The motivation for Axiom~\ref{sym2} is similar. Just as Axiom~\ref{sym1} expresses a form of aversion towards, Axiom~\ref{sym2} expresses a form of preference for uncertainty. 

\begin{proposition} \label{prop_sym}
Let $\succ$ admit a twofold multiprior representation $(u,C,D)$. Then:
\begin{enumerate}[\rm i)]
\item $\succ$ satisfies Axiom \ref{sym1} if and only if $D \subseteq C$. \label{prop_sym1}
\item $\succ$ satisfies Axiom \ref{sym2} if and only if $C \subseteq D$.
\label{prop_sym2}
\end{enumerate}
\end{proposition}

Proposition~\ref{prop_sym} states that under Axiom~\ref{sym1}, in order for the act $f$ to be preferred to different act $g$ (not necessarily complementary to $f$), $f$ must be evaluated against a larger set of probabilistic scenarios than $g$. The opposite conclusion holds under Axiom~\ref{sym2}. In particular, a concordant two-fold preference satisfies both axioms if and only if $C = D$, and thus the proposition provides a characterization of concordant twofold preferences.

\subsection{Comparative Uncertainty Aversion}
\label{sec_comparative}

We now derive formal comparisons of uncertainty attitudes across different agents. We follow the notion of comparative uncertainty aversion introduced by \cite*{ghirardato2002ambiguity}, adapted to our framework:

\begin{definition} \label{def_attitude}
Let $\succ_1$ and $\succ_2$ be two binary relations over $\calF$. We say that $\succ_1$ is \emph{more uncertainty averse over the alternative than} $\succ_2$ if for all $f \in \calF$ and $x \in X$,
\[
    f \succ_1 x \text{~~implies~~} f \succ_2 x.
\]
We say that $\succ_1$ is \emph{more uncertainty loving over the status-quo than} $\succ_2$ if for all $f \in \calF$ and $x \in X$,
\[
    x \succ_1 f \text{~~implies~~} x \succ_2 f.
\]
\end{definition}

An agent is more uncertainty averse than another if they are less prone to choosing an uncertain act $f$ over a constant status quo $x$. An agent is instead more uncertainty loving over the status quo if they are more inclined to stick to a status quo $f$ than to switch to an alternative $x$, even if the latter is a constant act. The next result characterizes comparative uncertainty attitudes for twofold preferences.

\begin{proposition} \label{prop_attitude}
Let $\succ_1$ and $\succ_2$ admit twofold multiprior representations $(u,C_1,D_1)$ and $(u,C_2,D_2)$, respectively. Then:
\begin{enumerate}[\rm i)]
\item $\succ_1$ is more uncertainty averse over the alternative than $\succ_2$ if and only if $C_2 \subseteq C_1$. \label{prop_attitude1}
\item $\succ_1$ is more uncertainty loving over the status quo than $\succ_2$ if and only if $D_2 \subseteq D_1$. \label{prop_attitude2}
\end{enumerate}
\end{proposition}

 In light of Proposition~\ref{prop_attitude}, one may view the set of beliefs  $C\setminus D$, used to evaluate the competing act, but not the status quo, as representing uncertainty-aversion tendencies. While the set $D\setminus C$ of beliefs that are used to evaluate the status quo, but not the alternative act, as capturing uncertainty-loving tendencies. 

 We may, in particular, provide a narrative for how agents with different sets of priors may differ in their attitudes toward uncertainty. Say that $A,B \subset \Delta(\Omega)$ are two disjoint, closed, and convex sets of priors. Consider the twofold multiprior preferences 
\[
    \begin{array}{cccc}
\succ_1 & \succ_2 & \succ_3 & \succ_4  \\
(u,A, A) & (u,\overline{\mathrm{co}}(A\cup B), A) & (u,A, \overline{\mathrm{co}}(A\cup B)) & (u,\overline{\mathrm{co}}(A\cup B), \overline{\mathrm{co}}(A\cup B))
\end{array}
\]
where $\overline{\mathrm{co}}$ denote the closed convex hull operator. As implied by Proposition~\ref{prop_attitude}, $\succ_2$ describes an agent whose preference is more uncertainty averse over the alternative than $\succ_1$, specifically, one that is reluctant to choose acts that are ambiguous in their evaluation through priors in $B$. Conversely, $\succ_3$ describes an agent who is more uncertainty loving over the status quo than $\succ_1$, one who is reluctant to choose a constant act over a status quo that could provide high expected utility under priors in $B$. Finally, $\succ_4$ is both more ambiguity averse and loving than $\succ_1$.

Next we relate a decision maker's attitude towards uncertainty to the degree of incompleteness of their preference relation:

\begin{definition}
Let $\succ_1$ and $\succ_2$ be two binary relations over $\calF$.
We say that $\succ_1$ is \emph{more conservative} than $\succ_2$ if for all $f,g \in \calF$,
\[
    f \succ_1 g \text{~~implies~~} f \succ_2 g.
\]
\end{definition}

Notice that if the relation $\succ_1$ is more conservative than the relation $\succ_2$, then by definition, $\succ_1$ must be both more uncertainty averse over the alternative and more uncertainty loving over the status quo than the $\succ_2$. For twofold multiprior preferences, Proposition \ref{prop_attitude} then implies $C_2 \subseteq C_1$ and $D_2 \subseteq D_1$. Moreover, the converse is true:

\begin{corollary} \label{cor_attitude}
Let $\succ_1$ and $\succ_2$ admit twofold multiprior representations $(u,C_1,D_1)$ and $(u,C_2,D_2)$, respectively. The preference relation $\succ_1$ is more conservative than $\succ_2$ if and only if $C_2 \subseteq C_1$ and $D_2 \subseteq D_1$.
\end{corollary}

\section{Connections to Other Decision Models} \label{sec_extension}

\subsection{Maxmin Expected Utility, Bewley, and Twofold Preferences}

In this section we flesh out the formal connection between subjective expected utility, Bewley preferences, and twofold multiprior preferences. The results are quite nuanced, but they imply that a preference relation is both a Bewley and a twofold multiprior preference if and only if it is consistent with subjective expected utility. Thus, the Bewley and twofold models are distinct alternatives to subjective expected utility.

In our discussion, we shall consider the relation between Axiom~\ref{axiom_contagion} and two other properties that are familiar in the literature:
 \begin{axiom}[Monotonicity]\label{ax:standard-monotonicity}
 For all $f,g \in \calF$, $f(\omega) \succ g(\omega)$ for all $\omega \in \Omega$ implies $f \succ g$.
 \end{axiom}
 \begin{axiom}[Independence]\label{ax:standard-independence}
 For all $f,g,h \in \calF$ and $\alpha \in (0,1)$, $f \succ g$ if and only if $\alpha f + (1 -\alpha) h \succ \alpha g + (1 - \alpha) h$.
 \end{axiom}
 The first axiom is the standard monotonicity axiom, while the second is the independence axiom of \cite{anscombe1963definition}. It is well known, and obvious, that subjective expected utility and Bewley's theory satisfy monotonicity and independence. Twofold preferences may not satisfy these properties.

 \begin{proposition}\label{thm:connections}
  Let $\succ$ be a binary relation on $\calF$.
  \begin{enumerate}[\rm i)]
  \item \label{it:conn1} If $\succ$ is a twofold multiprior preference, then it satisfies Axiom \ref{ax:standard-monotonicity} or \ref{ax:standard-independence} if and only if $\succ$ is a subjective expected utility preference.
  \item \label{it:conn2} If $\succ$ is a Bewley preference, then it satisfies Axiom \ref{axiom_contagion} if and only if $\succ$ is a subjective expected utility preference.
\end{enumerate}
\end{proposition}

The literature on the independence axiom is extensive. Suffice it to say here that there is plenty of empirical evidence on violations of independence, so it may be a desirable feature of twofold multiprior preferences that they do not impose this axiom.\footnote{While full independence is not satisfied, these preferences satisfy the more general ``dominance independence'' axiom introduced by \cite{faro2015variational}.\label{fn:faroreference}}

Monotonicity requires more of an explanation. Recall the interpretation behind our representation in terms of theories and evaluations. Alice's practical knowledge does not include a complete understanding of the state space, or the sets of theories that give rise to her set of evaluations. Her difficulty in comparing acts state-by-state makes her blind to the monotonicity expressed by the axiom. Finally, two-fold multiprior preferences are arguably in line with recent experimental evidence of violations of state-by-state dominance, and  monotonicity in different economic environments (see the references mentioned in the introduction).

Proposition~\ref{thm:connections} means that Bewley and twofold preferences describe very different methods for quantifying ambiguity. In Bewley's story, Alice has an exact state-by-state insight into each act. She is therefore able to make logical monotonicity and independence arguments when making decisions. Axiom~\ref{axiom_contagion} is rather unnecessary in the Bewley framework since constant acts are not unique in being fully understood. The incompleteness comes from a lack of knowledge about the ``right'' theory. For twofold preferences, in contrast, incompleteness arises because of the multiplicity of evaluations. 

\subsection{Who is More Conservative?}\label{sec:whoismoreconservative}

As in Section \ref{sec_comparative}, how conservatively the decision maker acts in the face of uncertainty can be expressed by the degree of incompleteness of her preference relation.

The next result shows how Bewley's preferences emerge as an extension of twofold multiprior preferences.
In other words, under the circumstances of the result we are about to present, a twofold conservative decision maker is more conservative than one in the corresponding Bewley model.

\begin{proposition}\label{thm:Bewley-Interval-Comparison}
Let $\succ$ be a twofold multiprior preference with representation $(u, C, D)$, and let $\succ^*$ be Bewley preference with representation $(u, C^*)$.
The preference relation $\succ$ is more conservative than $\succ^*$ if and only if $C^* \subseteq C \cap D$.
\end{proposition}

 Given such a preference $\succ$, it is always possible to find a Bewley preference that is less conservative, and the two preferences must be related by the inclusion $C^* \subseteq C \cap D$. The converse does not hold: Given a twofold preference $\succ$, one cannot find a non-trivial Bewley preference that is more conservative than $\succ$. This follows from Proposition~\ref{thm:connections}, which shows that $\succ$ does not satisfy the monotonicity axiom.

\subsection{Complete Extensions of Twofold Preferences}\label{sec:completeextensionsoftwofold}

 We now turn to the question of how to extend twofold multiprior preferences to complete preferences. Reasoning as in \cite*{gilboa2010objective}, here we think of the decision maker's decision process as described by two preference relations: $\succ$ and $\hat{\succ}$. The relation $\succ$ describes those preference judgements that the decision maker is able to make with sufficiently strong conviction. When expressing the ranking $f \succ g$, the decision maker can make a conclusive case, either trough evidence or logical arguments, that she is correct in her choices. Because of its demanding nature, this is a relation which is naturally incomplete. The preference $\hat{\succ}$ is an extension of $\succ$ (meaning that ${\succ}\subseteq{\hat{\succ}}$), and describes those preference rankings that the decision maker is forced to make, when making a choice. The relation $\hat{\succ}$ is assumed to be negatively transitive, i.e.\ the strict part of a complete preference relation.
 
 We depart from the existing literature by assuming that $\succ$ is a twofold multiprior preference. This means that $\succ$ describes choices that are not only well motivated, but also immediately convincing to agents who are unable to make state-by-state comparisons. We think of the ranking $f \succ g$ as saying that despite the decision maker's only partial understanding of the relation between states and consequences, she prefers $f$ to the status quo $g$. The preference $\succ^*$ is total binary relation that completes $\succ$. 

\begin{definition} \label{def_extension}
 Given two asymmetric binary relations $\succ$ and $\hat{\succ}$, we say that $\hat{\succ}$ is a \emph{compatible extension} of $\succ$ if ${\succ} \subseteq {\hat{\succ}}$ and the two coincide on $X$.
In particular, we say that $\hat{\succ}$ is a \emph{completion} of $\succ$ if it is a compatible extension of $\succ$ and satisfies negative transitivity.
\end{definition}

The main result of this section characterizes the class of completions that can be obtained from a given twofold multiprior preference $\succ$. First, we impose two weak continuity and monotonicity assumptions on $\hat{\succ}$. Specifically, we say that a completion $\hat{\succ}$ of $\succ$ is \emph{regular} if it satisfies the following conditions:

\begin{enumerate}[(R1).]
\item For all $f \in \calF$ and $x,y \in X$, the sets $\{\alpha \in [0,1]: \alpha x + (1-\alpha) y\ \hat{\succ}\ f\}$ and $\{\alpha \in [0,1]: f\ \hat{\succ}\ \alpha x + (1-\alpha) y\}$ are open. \label{regular1}
\item For all $f \in \calF$ and $x,y \in X$, if $x\ \hat{\succsim}\ f(\omega)\ \hat{\succsim}\ y$ for all $\omega \in \Omega$, then $f \not \hat{\succ}\ x$ and $y \not \hat{\succ}\ f$. \label{regular2}
\end{enumerate}

Relative to the axioms in Section~\ref{sec_main}, (R1) is slightly weaker than our continuity assumption, Axiom \ref{axiom_cts}, as it only involves mixtures between constant acts. Condition (R2) is in spirit similar to Axiom~\ref{axiom_mono}, but neither implies the other.

Specifically, (R2) requires that the negation of $\hat{\succ}$ satisfies the monotonicity axiom when at least one act involved in a choice is constant. Both conditions are permissive and satisfied by a broad class of preferences.\footnote{
Condition (R2) is automatically satisfied if the range of $u$ is an open subset of $\R$.
To see this, suppose that $x\ \hat{\succeq}\ f(\omega)\ \hat{\succeq}\ y$ holds for all $\omega$.
When $u(X)$ is open, we can perturb $x$ and $y$ in such a way that $x'\ \hat{\succ}\ f(\omega)\ \hat{\succ}\ y'$ holds for all $\omega$.
Since $\succ$ and $\hat{\succ}$ coincide on $X$, and since $\succ$ satisfies Axiom~\ref{axiom_mono}, it follows that $x' \succ f \succ y'$.
Moreover, since $x'$ and $y'$ can be arbitrarily close to $x$ and $y$ (in terms of the utility $u(\cdot)$ they produce), Axiom~\ref{axiom_cts} of $\succ$ implies $x \succ f \succ y$, from which $x\ \hat{\succ}\ f\ \hat{\succ}\ y$ obtains by the definition of completion.


On the other hand, condition (R1) is not necessarily derivable from the definition of completions.
We can construct lexicographic-type completions that systematically violate (R1), and thus, cannot be represented in the way described in Proposition \ref{prop_completion}. See Appendix \ref{appendix_example} for a concrete example.}

\begin{proposition} \label{prop_completion}
Let $\succ$ admit a twofold multiprior representation $(u,C,D)$, and let $\hat{\succ}$ be an asymmetric binary relation.
The following are equivalent:
\begin{enumerate}[\rm i)]
\item $\hat{\succ}$ is a regular completion of $\succ$.
\item There exists a function $\alpha\colon \calF \to [0,1]$ such that $\hat{\succ}$ is represented by:
\begin{align}
V(f) = (1-\alpha (f)) \min_{\mu \in C} \int u(f) \d\mu + \alpha (f) \max_{\mu \in D} \int u(f) \d\mu. \label{eq_alpha}
\end{align}
\end{enumerate}
\end{proposition}

 We refer to preferences that have a representation of the form \eqref{eq_alpha} as \emph{generalized $\alpha$-maxmin preferences}. This representation is general in that the function $\alpha (\cdot)$ can vary across different acts, and that the sets of priors $C$ and $D$ can differ. In particular, these properties allow for the possibility that $\hat{\succ}$ violates independence or monotonicity. Some important special cases are \emph{$\alpha$-maxmin preferences}, where $C = D$ and $\alpha$ is constant \citep*{hurwicz1951some,frick2021objective}, as well as \emph{invariant biseparable preferences} \citep*{ghirardato2004differentiating}, where $C = D$ and $\alpha$ satisfies an additional measurability condition.

 Motivated by Proposition~\ref{prop_completion} on generalized $\alpha$-maxmin preferences, we can study to the case of maxmin, in which $\alpha=0$, and maxmax, in which $\alpha=1$. The notion of caution, defined below, captures a subjective desire to prefer constant acts over uncertain acts, and is due to \cite*{gilboa2010objective}. The notion of abandon captures a decision maker who prefers uncertain acts over constant acts. The two axioms characterize maxmin and maxmax preferences.

\begin{definition}
We say that two binary relations $\succ$ and $\hat{\succ}$ jointly satisfy \emph{caution} if for all $f\in \calF$ and $x \in X$, $f \not \succ x$ implies $f \not \hat{\succ}\ x$.
Also, $\succ$ and $\hat{\succ}$ jointly satisfy \emph{abandon} if for all $f \in \calF$ and $x \in X$, $x \not \succ f$ implies $x \not \hat{\succ}\ f$.
\end{definition}

\begin{corollary} \label{cor_maximin}
Let $\succ$ admit a twofold multiprior representation $(u,C,D)$, and let $\hat{\succ}$ be a regular completion of $\succ$.
A pair $(\succ,\hat{\succ})$ jointly satisfies caution (resp.\ abandon) if and only if $\hat{\succ}$ admits the maxmin (resp.\ maxmax) representation by $(u,C)$ (resp.\ $(u,D)$).
\end{corollary}

\section{Empirical Content of Twofold Preferences} \label{sec_rational}

\subsection{Weakly Rationalizable Choice}
\label{sec:choice}

We now turn to the positive empirical content of twofold preferences for observable choices. Section~\ref{sec_main} provides an axiomatic foundation for twofold preferences, but it may be difficult to interpret it as a positive foundation. Here we consider instead an explicit model of choice under a status quo.

In particular, we show that twofold preferences can rationalize strictly more choice behaviors than Bewley's model. Every choice function that is consistent with the Bewley model is also consistent with our model of twofold preferences; and there are some choice behaviors that are consistent with the twofold model but not with Bewley's. 

We start by formulating model of choice that accommodates preference incompleteness. Let $\calA$ be a collection of finite sets of acts, which we call \emph{menus}. A \emph{choice function} is a function $c\colon \calA \times \calF \to \calF$ that maps each pair of a menu $A \in \calA$ and a status quo act $h \in \calF$ into the choice $c(A;h) \in A \cup \{h\}$.
We are interested in choice behavior when some acts may not be comparable, which translates into a bias towards choosing a status-quo act.

Given a binary relation $\succ$ on $\calF$, we say that a choice function $c$ is \emph{weakly rationalizable by $\succ$ under the status quo $h$} if for all $A \in \calA$, we have
\begin{itemize}
\item $c(A;h)=h$ implies $f \not \succ h$ for all $f \in A$; and
\item $c(A;h)=f\neq h$ implies that $f \succ h$, and $g \not \succ f$ for all $g \in A$.\footnote{
Here weak rationalizability stands in contrast with strong rationalizability, which demands that the chosen element is the unique maximal element according to the binary relation. See \cite{chambers2016revealed} for a discussion of this distinction.}
\end{itemize}

The next proposition shows that when one restricts attention to choice problems with a constant status quo, twofold preferences can empirically explain a strictly wider array of agent choice functions than Bewley preferences.

\begin{proposition}\label{thm:choice}
Let $x \in X$ be any constant act that serves as the status quo.
If a choice function is weakly rationalizable under $x$ by a Bewley preference with representation $(u,C)$, then it is weakly rationalizable under $x$ by a twofold multiprior preference with representation $(u,C,C)$.
If $u(x) < \sup u(X)$, however, there is a choice function that is weakly rationalizable under $x$ by a twofold multiprior preference but not by any Bewley preference.
\end{proposition}

Proposition~\ref{thm:choice} suggests that twofold preferences need not entail much loss of descriptive viability. It also shows that twofold preferences with $C \neq D$ can explain choice behaviors that cannot be captured through either Bewley's model, or concordant twofold preferences. Intuitively, this reflects the ability of non-concordant preferences to capture uncertainty aversion and love at the same time, as shown in Proposition~\ref{prop_attitude}.

\subsection{Undominated Choice: An Application to Second-price Auctions} \label{sec_auctions}

 We apply our notion of representation to the classical setting of sealed-bid second-price auctions. Our application demonstrates that twofold preferences are able to explain a wide array of bidding behaviors, yet they can provide sharper predictions than obvious dominance of \cite{li}.

 Consider a single indivisible good to be auctioned off. We fix an arbitrary buyer, referred to as Alice, and study her bidding behavior. Denote a generic outcome for our decision maker by $(x,t) \in [0,1] \times \R$, where $x$ is the chance of winning the good and $t$ is her payment to the auctioneer.
 Assume that Alice's utility function takes the quasi-linear form $u(x,t)=vx-t$, where $v \in V = \{0, 1, \dots, \bar{b}\}$ stands for her true valuation of the good.
 The value $v$ is fixed throughout this section.\footnote{We discretize the set of valuations to keep it consistent with the finite state space in our decision-theoretic analysis, but all results in this section are easy to extend to the continuum setting. One minor difference is that truth-telling in the continuum setting becomes a unique dominant strategy in the standard state-wise sense, but the same uniqueness is not guaranteed in the discrete setting; given that ties are broken in favor of Alice, she is indifferent between reporting $v$ and $v-1$, regardless of opponent's bid.}

 We model the uncertainty faced by Alice through the state space $\Omega = \{0, 1, \dots, \bar{b}\}$, where $\omega \in \Omega$ expresses the highest bid among opponents, which summarizes all payoff-relevant uncertainty. 
 When Alice is endowed with a single prior $\mu \in \Delta(\Omega)$, her own bid $b$ in the second-price auction induces the subjective expected utility defined as
\begin{align*}
U(b, \mu)  = \int_0^b (v-\omega) \d \mu(\omega).
\end{align*}
Here, we assume that ties are broken in favor of Alice for simplicity.
A prominent feature of the second-price auction is that the truth-telling strategy, i.e., $b = v$ maximizes $U(\cdot,\mu)$ for any prior $\mu$.

 When Alice deals with uncertainty through multiprior there can exist a range of strategies other than truth-telling that are not dominated.
 Specifically, assume that she conceives non-disjoint, closed, and convex sets $C, D \subseteq \Delta(\Omega)$ of priors, and that she evaluates each strategy based on the corresponding twofold multiprior representation.
 Then, the set of undominated bidding strategies is defined by
\begin{align} \label{undom-bids}
B^*(C, D) = \left\{
b \in V: \not \exists b' \in V \ \text{ s.t. }
\min_{\mu \in C} U(b',\mu)
> \max_{\mu \in D} U(b,\mu) \right\}.
\end{align}

The above definition generalizes obvious dominance by allowing for more flexibility in belief sets. Compared with the model of weak rationalizability studied in the previous section, undominated choice does not rely on the existence of a status-quo option. Yet, we can formally relate these two notions as follows: Consider any choice function $c(V;b)$, associated with prior sets $C$ and $D$, over the grand set $V$ of valuations.
Here, we view status-quo strategy $b$ as a variable. One can easily confirm that
\begin{align*}
B^*(C,D) = \{b \in V: c(V;b) = b\}.
\end{align*}
Namely, the set of undominated bids coincides with the set of those $b$ that are chosen by the weakly rationalizable choice function when Alice considers $b$ as the status-quo strategy.

The next proposition reveals the structure of undominated bids.
Specifically, the set of undomoinated bids will be given as a discrete interval that contains the true valuation $v$.

\begin{proposition} \label{prop_auc1}
There exist $b_*, b^* \in V$ with $b_* \leq v \leq b^*$ such that $B^*(C, D)=\{b_*, \dots, b^*\}$.
\end{proposition}

\begin{figure}[t!]
\centering
\includegraphics[width=11cm]{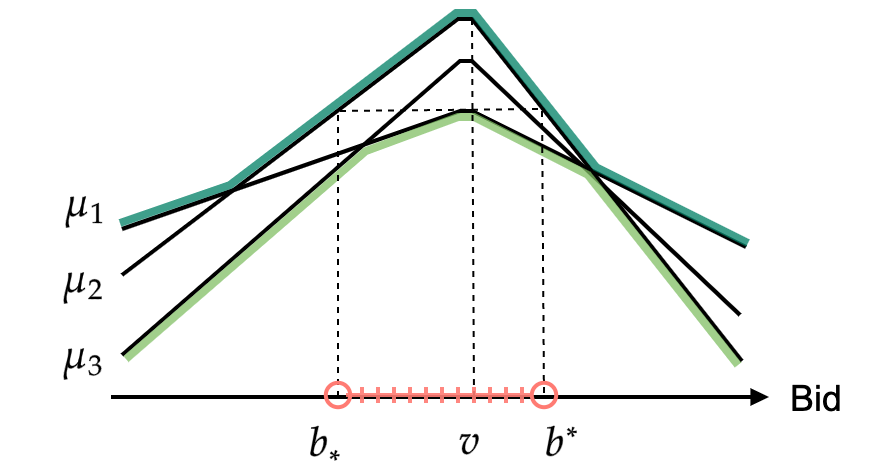}
\caption{The illustration of Proposition \ref{prop_auc1}.}
\label{fig_auction}
\end{figure}

The intuition behind Proposition~\ref{prop_auc1} can be explained using Figure~\ref{fig_auction}.
Consider a decision maker who holds three priors, say $\mu_1$, $\mu_2$, and $\mu_3$, according to which the expected utility is computed as functions of $b$.
While these expected utility functions vary across priors, all of them share the same properties: They are weakly increasing on $\{0, \dots ,v\}$, weakly decreasing on $\{v, \dots, \bar{b}\}$, and achieves a maximum at $b=v$.
More importantly, notice that the same properties will be satisfied by the lower and upper envelops of expected utility functions.
This implies that whether a given $b$ belongs to the undominated set can be determined by checking if the value of the upper envelope at $b$ is no less than the value of the lower envelope at $v$.
As is depicted in the graph, the single-peakedness implies the region of such $b$ takes the form of a discrete interval that certainly contains $v$.

 When all possible beliefs are considered relevant, only a trivial set of predictions is available.

\begin{proposition}\label{obvious}
If $C=D=\Delta(\Omega)$, then $B^*(C, D) = \{0, \dots,\bar{b}\}$.
\end{proposition}

Finally, our comparative statics results (Proposition \ref{prop_attitude} and Corollary \ref{cor_attitude}) are readily translated to the present context. Namely, the set of undominated bids shrinks as the sets of priors get smaller.

\begin{corollary}\label{coro_auc2}
If $C_2 \subseteq C_1$ and $D_2 \subseteq D_1$, then $B^*(C_2, D_2) \subseteq B^*(C_1, D_1)$.
\end{corollary}

\section{Twofold Maxmin Representation}
\label{sec_maxmin}

 There are two ways in which a twofold multiprior representation expresses a propensity to choose the status quo. First, by allowing for an asymmetric evaluation when comparing $f$ versus $g$ or $g$ versus $f$, the representation departs from the completeness axiom. Second, by applying a worst-case criterion to the alternative, but a best-case criterion to the status quo, the representation further amplifies the decision maker's inclination to choose the status-quo option. This second feature is crucial and it is behaviorally characterized by Axiom~\ref{axiom_conv} on the convexity of contour sets.
 
 While convexity of upper-contour sets is a standard assumption for modeling uncertainty aversion, convexity of lower-contour sets is a less common condition. Moreover, the normative rationale for the asymmetry in ambiguity attitude between the status quo and the alternative is not entirely self-evident.
 
 In this section we study an alternative representation that displays uncertainty aversion with respect to both the alternative and the status quo, while maintaining the general structure of twofold preferences:

\begin{definition}
 A binary relation $\succ$ is a \emph{twofold maxmin preference} if there exist a non-constant affine function $u\colon X \to \R$, and two non-empty closed convex subsets $D \subseteq C \subseteq \Delta (\Omega)$ such that for all $f,g \in \calF$,
\begin{align} \label{eq_maxmin}
 f \succ g \Longleftrightarrow \min_{\mu \in C} \int u(f) \d \mu >  \min_{\mu \in D} \int u(g) \d \mu.
\end{align}
\end{definition}

 Both the twofold multiprior and the twofold maxmin representations display more uncertainty aversion in the evaluation of the alternative $f$ than the in the evaluation of the status quo $g$, albeit in different ways. In the multiprior representation, $f$ is evaluated according to a uncertainty averse criterion, and $g$ according to a uncertainty loving criterion. In the twofold maxmin representation, both options are instead evaluated according to a maxmin criterion, but the evaluation applied to the alternative $f$ displays more uncertainty aversion than the one applied to the status quo. This follows from the hypothesis that $D$ is a subset of $C$.

\begin{figure}[t!]
\begin{center}
\includegraphics[width=15cm]{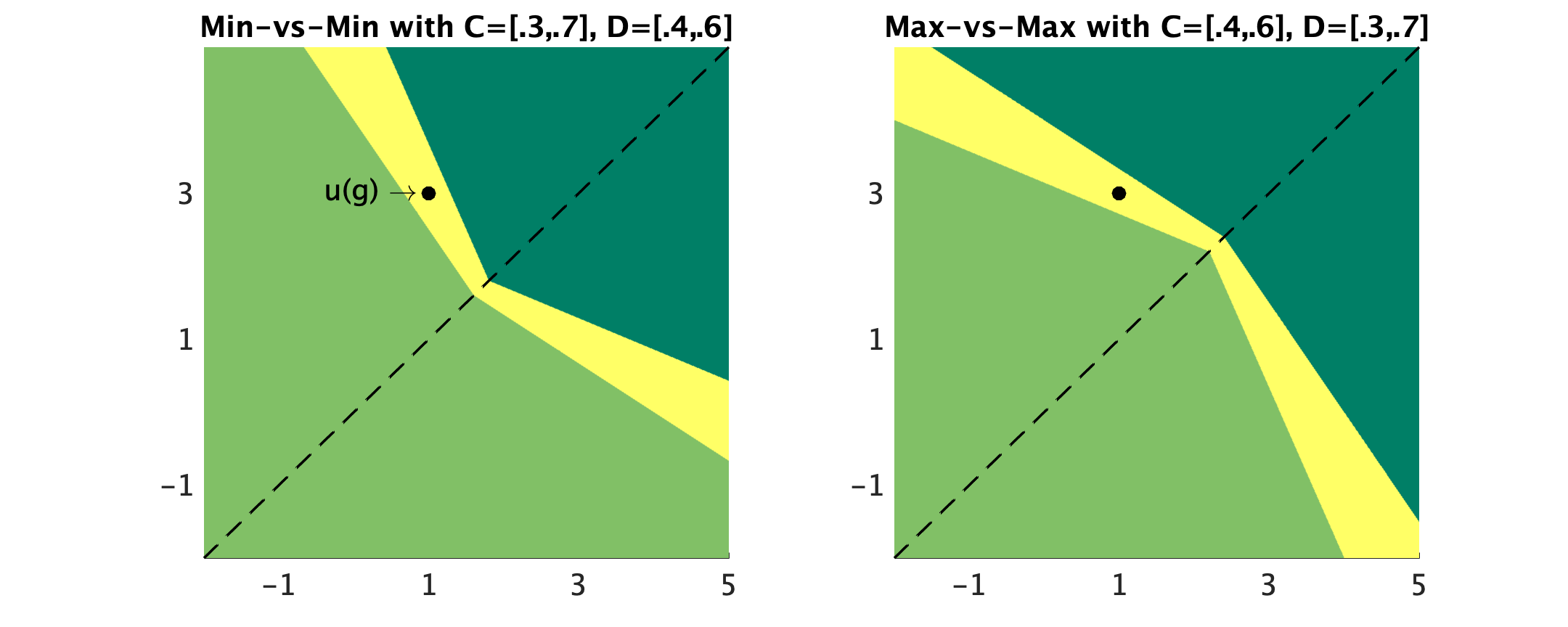}
\caption{Upper contour sets (dark green), lower contour sets (light green), and incomparable sets (yellow) are displayed for each of twofold maxmin (left) and maxmax (right) preferences.
The status-quo point $u(g)$ is fixed to be $(1,3)$, displayed as black bullets.
All boundaries are included in the yellow regions by the openness of $\succ$.
}
\label{fig_contour3}
\end{center}
\end{figure}

 Note that when $C=D$, the representation (\ref{eq_maxmin}) reduces to the standard maxmin representation of \cite{gilboa1989maxmin}. In terms of contour sets, as in the left panel of Figure \ref{fig_contour3}, the twofold maxmin preference is featured by convexity of the set of acts that do not dominate the status quo (i.e.\ convexity of the union of yellow and dark green regions). This graphical observation guides us to consider the following alternative of Axiom~\ref{axiom_conv}.

\begin{axiom} \label{axiom_averse}
For all $x \in X$, the sets $\{f \in \calF: f \succ x\}$ and $\{f \in \calF: x \not \succ f\}$ are convex.
\end{axiom}

The convexity of $\{f \in \calF: f \succ x\}$ implies that if both $f$ and $g$ were already understood as strictly better than $x$, then their mixture $\alpha f + (1-\alpha) g$ should also be ranked better than $x$. On the other hand, the convexity of $\{f \in \calF: x \not \succ f\}$ implies that if both $f$ and $g$ were, at least, not worse than $x$, then the mixture $\alpha f + (1-\alpha) g$ should not be worse than $x$ as well. Therefore, Axiom~\ref{axiom_averse} requires that hedging weakly improves the perceived quality of uncertain acts, both with respect to status-quo acts and alternative acts.

On top of the other axioms introduced in Section \ref{sec_main}, the next theorem shows that Axiom~\ref{axiom_averse} is necessary and sufficient for twofold maxmin preferences.

\begin{theorem} \label{thm_maxmin}
A binary relation $\succ$ satisfies Axioms~\ref{axiom_str}--\ref{axiom_cindp}, \ref{axiom_contagion}--\ref{axiom_mono}, and \ref{axiom_averse} if and only if it admits a twofold maxmin representation.
Moreover, $u$ is unique up to positive affine transformations, and $C$ and $D$ are unique.
\end{theorem}

We can also consider the converse of Axiom~\ref{axiom_averse} by postulating the convexity of lower contour sets. Given the results of Theorem~\ref{thm_maxmin}, it may not be surprising that Axiom~\ref{axiom_loving} creates the decision maker's uncertainty loving attitudes with respect to both alternative and status quo.

\begin{axiom} \label{axiom_loving}
For all $x \in X$, the sets $\{f \in \calF: x \succ f\}$ and $\{f \in \calF: f \not \succ x\}$ are convex.
\end{axiom}

Analogously to twofold maxmin, one can define a \emph{twofold maxmax representation} by replacing the min-operators in both sides of (\ref{eq_maxmin}) with the max-operators, and by replacing the condition $D \subseteq C$ with $C \subseteq D$.
The right panel of Figure \ref{fig_contour3} illustrates the contour sets generated by such preferences.
The proof of the next theorem is essentially the same as the preceding one, and hence we omit it.

\begin{theorem} \label{thm_maxmax}
A binary relation $\succ$ satisfies Axioms~\ref{axiom_str}--\ref{axiom_cindp}, \ref{axiom_contagion}--\ref{axiom_mono}, and \ref{axiom_loving} if and only if it admits a twofold maxmax representation.
Moreover, $u$ is unique up to positive affine transformations, and $C$ and $D$ are unique.
\end{theorem}

 From a modeling point of view, the twofold multiprior representation can be seen as one special case of a general approach to model incomplete preferences by applying different representations for upper and lower contour sets. In this regard, our auxiliary result (Lemma~\ref{lem_rep}) in Appendix~\ref{app_pf_thm} demonstrates that a more general form of twofold representation can be derived even without Axiom~\ref{axiom_conv}. The follow-up work by \cite{cusumano2021twofold} identifies the exact set of axioms that are necessary and sufficient for general twofold representation. They also show that the twofold approach is not limited to what is considered in the present paper, but is also compatible with several weakening of independence-type axioms, such as \emph{comonotonic independence} by \cite{schmeidler1989subjective} and \emph{weak C-independence} by \cite{maccheroni2006ambiguity}.

\newpage

\renewcommand{\thesection}{\Alph{section}}
\setcounter{section}{0}
\section{Appendix: Omitted Proofs}

\renewcommand{\thesubsection}{A.\arabic{subsection}}

\subsection{Proof of Theorem \ref{thm_main}}
\label{app_pf_thm}

Throughout the appendix, we denote by $\cl$, $\co$, and $\cco$ to express the closure, convex hull, and closed convex hull operators, respectively.
We will work with the finite-dimensional space $\R^\Omega$, equipped with the sup-norm $\|\xi\| = \max_{\omega \in \Omega} |\xi(\omega)|$.
Given any number $r \in \R$, let $r\1_\Omega$ denote a constant vector that takes $r$ for every $\omega$-th coordinate, and let $\R^\Omega_{\rm diag} = \{r\1_\Omega \in \R^\Omega: r \in \R\}$.
As is standard, let $\R_+^\Omega = \{\xi \in \R^\Omega: \xi(\omega) \ge 0,\, \forall \omega \in \Omega \}$ and $\R_{++}^\Omega = \{\xi \in \R^\Omega: \xi(\omega) > 0,\, \forall \omega \in \Omega \}$, respectively. The sets $\R^\Omega_-$ and $\R^\Omega_{--}$ are defined analogously.

\label{appendix_main}

\subsubsection{Sufficiency}
We first prove the sufficiency of the axioms.
Let $\succ$ be a binary relation that satisfies Axioms \ref{axiom_str}--\ref{axiom_mono}. (We remark that Axiom \ref{axiom_conv} will no be used until Claim \ref{claim_functional}.)
Consider the restriction of $\succ$ to constant acts.
Recall that the binary relation $\succsim$ on $X$ is defined, for all $x,y \in X$, by $x \succsim y$ if and only if $y \not \succ x$.
Since $\succ$ is asymmetric and negatively transitive on $X$ by Axiom \ref{axiom_str}, $\succsim$ is complete and transitive.
Together Axioms \ref{axiom_cts} and \ref{axiom_cindp}, it then follows that $\succsim$ satisfies all the von Neumann-Morgenstern axioms.
Hence, there exists an affine function $u\colon X \to \R$ such that $x \succsim y$ if and only if $u(x) \ge u(y)$ for all $x,y \in X$.
In particular, $u$ is non-constant since $\succ$ is non-trivial, and $u$ is unique up to positive affine transformations.

\begin{claim} \label{claim_nonemp}
For all $f \in \calF$, the set $\{x \in X: f \bowtie x\}$ is non-empty.
\end{claim}

\begin{proof}
Fix $f \in \calF$. Since $\Omega$ is finite, we can find $\overline{x}, \underline{x} \in f(\Omega)$ such that $\overline{x} \succsim f(\omega) \succsim \underline{x}$ for all $\omega \in \Omega$.
We claim that $f \not \succ \overline{x}$ and $\underline{x} \not \succ f$.
First consider the case where $\overline{x} \sim \underline{x}$, then $f(\omega) \sim \overline{x}$ for all $\omega \in \Omega$.
Since $\overline{x} \not \succ \overline{x}$ by asymmetry, in this case, Axiom \ref{axiom_equiv} implies that $f \not \succ \overline{x}$.
Similarly, we have $\underline{x} \not \succ f$ by Axiom \ref{axiom_equiv} and the fact that $f(\omega) \sim \underline{x}$ for all $\omega \in \Omega$.
Now consider the case $\overline{x} \succ \underline{x}$, and suppose by way of contradiction that $f \succ \overline{x}$.
By Axiom \ref{axiom_cts}, we can find $\alpha \in (0,1)$ such that $g \equiv \alpha f + (1-\alpha) \underline{x} \succ \overline{x}$.
However, since $u$ is affine, it follows that
\[
u(g(\omega)) = \alpha u (f(\omega)) + (1-\alpha) u(\underline{x}) < u(\overline{x}),
\]
for all $\omega \in \Omega$, so Axiom \ref{axiom_mono} dictates $\overline{x} \succ g$, a contradiction.
Hence, it follows that $f \not \succ \overline{x}$, and we can similarly show that $\underline{x} \not \succ f$ as well.

Given any $\lambda \in [0,1]$, let $x_\lambda = \lambda \overline{x} + (1-\lambda) \underline{x}$.
We define the sets $\Lambda, \Lambda' \subseteq [0,1]$ by
\begin{alignat*}{2}
\Lambda = \{\lambda \in [0,1]: f \not \succ x_\lambda\} \ \text{ and } \ 
\Lambda' = \{\lambda \in [0,1]: x_\lambda \not \succ f\}.
\end{alignat*}
Both $\Lambda$ and $\Lambda'$ are non-empty by the preceding arguments, and closed relative to $[0,1]$ by Axiom \ref{axiom_cts}.
Moreover, we have $\Lambda \cup \Lambda' = [0,1]$.
Indeed, if $\lambda \in [0,1] \setminus \Lambda$ then we have $f \succ x_\lambda$, so the asymmetry of $\succ$ yields $x_\lambda \not \succ f$.
Hence, we have $\lambda \in \Lambda'$ whenever $\lambda' \not\in \Lambda$, from which $\Lambda \cup \Lambda' = [0,1]$.
Thus, the connectedness of $[0,1]$ implies that $\Lambda \cap \Lambda' \neq \emptyset$, and therefore, there exists some $\lambda^* \in \Lambda \cap \Lambda'$.
By construction, it follows that $f \bowtie x_{\lambda^*}$ as desired.
\end{proof}

As is discussed in Section \ref{sec_main}, we define two functions $\overline{U}, \underline{U}\colon \calF \to \R$ as the values of {\it maximal} and {\it minimal} constant acts that are not comparable with a given act.
Formally, we set
\begin{align}
\overline{U}(f) = \sup_{x \in X} \{u(x): f \bowtie x\} \ \text{ and } \ 
\underline{U}(f) = \inf_{x \in X} \{u(x): f \bowtie x \}, \tag{\ref{eq_def_util}}
\end{align}
for every $f \in \calF$. Claim \ref{claim_nonemp} ensures that the set $\{u(x): f \bowtie x\}$ is nonempty.
Moreover, letting $\overline{x}$ and $\underline{x}$ respectively denote the best and worst outcomes that result from $f$, Axiom \ref{axiom_mono} leads to $u(\overline{x}) \ge \overline{U}(f)$ and $\underline{U}(f) \ge u(\underline{x})$, so both functions are real-valued.
Let us collect a few other properties in the next claim.

\begin{claim} \label{claim_property}
The functions $\underline{U}, \overline{U}\colon \calF \to \R$ possess the following properties.
\begin{enumerate}[\rm i)]
\item For all $x \in \calF$, $u(x) = \overline{U}(x) = \underline{U}(x)$. \label{claim_property1}
\item For all $f \in \calF$, $u(\overline{x}) \ge \overline{U}(f) \ge \underline{U}(f) \ge u(\underline{x})$, where $\overline{x}, \underline{x} \in f(\Omega)$ are outcomes such that $u(\overline{x}) \succsim f(\omega) \succsim u(\underline{x})$ for all $\omega \in \Omega$. \label{claim_property2}
\item $u(X) = \underline{U}(\calF) = \overline{U}(\calF)$.
\label{claim_property3}
\end{enumerate}
\end{claim}

\begin{proof}
Clearly, (\ref{claim_property1}) follows from the fact that $\succ$ is asymmetric and negatively transitive, and thus, constitutes the strict part of a weak order.
As for (\ref{claim_property2}), fix any $f \in \calF$, and let $\overline{x}$ and $\underline{x}$ be as in the statement.
Clearly, $\overline{U}(f) \ge \underline{U}(f)$ holds by construction.
Also, $u(\overline{x}) \ge \overline{U}(f)$ is trivially true if $u(\overline{x}) = \sup u(X)$.
So, assume that $u(\overline{x}) < \sup u(X)$, and consider any small $\epsilon > 0$ and $x_\epsilon \in X$ such that $u(\overline{x}) < u(x_\epsilon) \le u(\overline{x}) + \epsilon$.
It holds that $x_\epsilon \succ \overline{x} \succsim f(\omega)$ for all $\omega \in \Omega$, so Axiom \ref{axiom_mono} implies $x_\epsilon \succ f$.
Now, if $u(x_\epsilon) < \underline{U}(f)$, then we can find $y \in X$ with $f \bowtie y$ with $u(x_\epsilon) < u(y)$.
But then, transitivity dictates $y \succ f$, a contradiction.
Hence, it follows that $u(\overline{x}) + \epsilon \ge u(x_\epsilon) > \overline{U}(f)$, and thus, letting $\epsilon \to 0$ yields $u(\overline{x}) \ge \overline{U}(f)$.
The proof of $\underline{U}(f) \ge u(\underline{x})$ is symmetric.
As for (\ref{claim_property3}), note that (\ref{claim_property1}) readily implies $u(X) \subseteq \underline{U}(\calF)$ and $u(X) \subseteq \overline{U}(\calF)$.
Moreover, since $u(X)$ is convex by the affinity of $u$, the inequalities $u(\overline{x}) \ge \overline{U}(f) \ge \underline{U}(f) \ge u(\underline{x})$ imply the converse inclusions as well.
\end{proof}

The next lemma derives an important implication of Axiom \ref{axiom_contagion}.

\begin{lemma} \label{lem_calibration}
In the presence of Axioms \ref{axiom_str}--\ref{axiom_cindp}, \ref{axiom_equiv}, and \ref{axiom_mono}, a binary relation $\succ$ satisfies Axiom \ref{axiom_contagion} if and only if for all $f,g \in \calF$ with $f \succ g$ there exists some $x,y \in X$ such that $f \succ x \succ y \succ g$.
\end{lemma}

\begin{proof}
We first show how the stated property implies Axiom~\ref{axiom_mono}. Consider a pair of comparable acts $f \succ g$.
If the stated property is satisfied, then we can take $x,y \in X$ such that $f \succ x \succ y \succ g$.
Now observe that any $z \in X$ must be comparable with either $x$ or $y$, so there are four possibilities that we have to consider:\ $x \succ z$, $z \succ x$, $y \succ z$, or $z \succ y$.
In each of these cases, since $\succ$ is transitive, one can easily check that $z$ is comparable with either $f$ or $g$.
Hence, $\succ$ satisfies Axiom \ref{axiom_contagion}.

Conversely, suppose that $\succ$ satisfies Axiom \ref{axiom_contagion}, and fix any $f,g \in \calF$ with $f \succ g$.
Since $\Omega$ is finite, we can find $\succsim$-maximal and $\succsim$-minimal elements from $f(\Omega) \cup g(\Omega)$, which are denoted by $\overline{x}$ and $\underline{x}$.
If $\overline{x} \sim \underline{x}$, then $f(\omega) \sim g(\omega)$ for all $\omega \in \Omega$.
But, since $f \succ g$, Axiom \ref{axiom_equiv} implies $f \succ f$, a contradiction to asymmetry.
Hence, $\overline{x} \succ \underline{x}$ holds.
We set
\begin{align*}
f' = \frac{1}{2} f + \frac{1}{4} \overline{x} + \frac{1}{4} \underline{x} \ \text{ and } \ g' = \frac{1}{2} g + \frac{1}{4} \overline{x} + \frac{1}{4} \underline{x}.
\end{align*}
Axiom \ref{axiom_cindp} implies that $f' \succ g'$.
Moreover, for any $\omega \in \Omega$, the affinity of $u$ implies that $u(\overline{x}) > u(f'(\omega))$ and $u(g'(\omega)) > u(\underline{x})$, from which Axiom \ref{axiom_mono} yields $\overline{x} \succ f'$ and $g' \succ \underline{x}$.
Summarizing these relations using transitivity, we have $\overline{x} \succ f' \succ g' \succ \underline{x}$.
Now, given any $\lambda \in [0,1]$, let $x_\lambda = \lambda \overline{x} + (1-\lambda) \underline{x}$. 
We define the sets $\Lambda, \Lambda' \subseteq [0,1]$ by
\begin{alignat*}{2}
\Lambda = \{\lambda \in [0,1]: f' \succ x_\lambda\} \ \text{ and } \ 
\Lambda' = \{\lambda \in [0,1]: x_\lambda \succ g'\}.
\end{alignat*}
Note that both are non-empty since $0 \in \Lambda$ and $1 \in \Lambda'$, and open relative to $[0,1]$ by Axiom \ref{axiom_cts}.
Moreover, we claim that $\Lambda \cup \Lambda' = [0,1]$.
To see this, let $\lambda \in [0,1] \setminus \Lambda$, that is, $f' \not \succ x_{\lambda}$.
If $x_{\lambda} \succ f'$, then transitivity implies $x_{\lambda} \succ g'$, so $\lambda \in \Lambda'$.
Otherwise, we have $f' \bowtie x_{\lambda}$, but since $f' \succ g'$, Axiom \ref{axiom_contagion} implies that $x_{\lambda}$ must be comparable with $g'$.
Namely, either $x_{\lambda} \succ g'$ or $g' \succ x'_{\lambda}$ holds.
In the latter case, transitivity yields $f' \succ x'_{\lambda}$, a contradiction to $f' \bowtie x'$.
Hence, we must have $x_{\lambda} \succ g'$, thereby $\lambda \in \Lambda'$.
Therefore, the connectedness of $[0,1]$ implies $\Lambda \cap \Lambda' \neq \emptyset$.
In particular, since $\Lambda \cap \Lambda'$ is open, we can take distinct numbers $\lambda_1 > \lambda_2$ from this set.
By construction, if we write $x_1 = x_{\lambda_1}$ and $x_2 = x_{\lambda_2}$, it follows that $\overline{x} \succ f' \succ x_1 \succ x_2 \succ g' \succ \underline{x}$.

The affinity of $u$ lead to $\frac{3}{4} \overline{x} + \frac{1}{4} \underline{x} \succ f'(\omega)$ and $g'(\omega) \succ \frac{1}{4} \overline{x} + \frac{3}{4} \underline{x}$ for all $\omega \in \Omega$.
Hence, Axiom \ref{axiom_mono} implies that
\begin{align*}
\frac{3}{4} \overline{x} + \frac{1}{4} \underline{x}
\succ f'
\succ x_1 \succ x_2
\succ g'
\succ \frac{1}{4} \overline{x} + \frac{3}{4} \underline{x},
\end{align*}
Moreover, by Axiom \ref{axiom_cts}, we can find $\alpha_1,\alpha_2 \in [\frac{1}{4}, \frac{3}{4}]$ with $\alpha_1 > \alpha_2$ such that $x_i \sim \alpha_i \overline{x} + (1-\alpha_i) \underline{x}$ for $i \in \{1,2\}$.
Equivalently, we can find $\beta_1,\beta_2 \in [0,1]$ with $\beta_1 > \beta_2$ such that $x_i \sim \frac{1}{2}(\beta_i \overline{x} + (1-\beta_i) \underline{x}) + \frac{1}{4}\overline{x} + \frac{1}{4} \underline{x}$ for $i \in \{1,2\}$.
Letting $y_i = \beta_i \overline{x} + (1-\beta_i) \underline{x}$, we obtain
\begin{align*}
f' = \frac{1}{2}f + \frac{1}{4} \overline{x} + \frac{1}{4} \underline{x}
\succ \frac{1}{2}y_1 + \frac{1}{4} \overline{x} + \frac{1}{4} \underline{x}
\succ \frac{1}{2}y_2 + \frac{1}{4} \overline{x} + \frac{1}{4} \underline{x}
\succ \frac{1}{2}f + \frac{1}{4} \overline{x} + \frac{1}{4} \underline{x} = g'.
 \end{align*}
 Axiom \ref{axiom_cts} now implies $f \succ y_1 \succ y_2 \succ g$ as desired.
\end{proof}

We are now ready to establish the representation of $\succ$ that employs two utility functions $\underline{U}, \overline{U}\colon \calF \to \R$.

\begin{lemma} \label{lem_rep}
Suppose that $\succ$ satisfies Axioms \ref{axiom_str}--\ref{axiom_cindp} and \ref{axiom_contagion}--\ref{axiom_mono}, and let $\underline{U}, \overline{U}\colon \calF \to \R$ be defined as in (\ref{eq_def_util}).
For all $f,g \in \calF$, $f \succ g$ if and only if $\underline{U}(f) > \overline{U} (g)$.
\end{lemma}

\begin{proof}
Suppose $\underline{U}(f) > \overline{U} (g)$. In proving that $f \succ g$ we first consider the case where either $f$ or $g$ is a constant act.
So suppose that $\underline{U}(f) > u(x)$.
By the definition of $\underline{U}$, $f$ and $x$ must be comparable, i.e., either $f \succ x$ or $x \succ f$.
In particular, for any $y \in X$ with $f \bowtie y$, we have $u(y) \ge \underline{U}(f) > u(x)$, so $y \succ x$ holds.
Hence, we cannot have $x \succ f$ in the presence of transitivity.
So, $f \succ x$ holds whenever $\underline{U}(f) > u(x)$.
Similarly, we can show that $x \succ g$ holds whenever $u(x) > \overline{U}(g)$.
Now, we move on to the general case.
For $f,g \in \calF$, suppose that $\underline{U}(f) > \overline{U} (g)$.
Note that the values $\underline{U}(f)$ and $\overline{U}(g)$ belong to the set $u(X)$, and hence, we can pick some $x \in X$ such that $\underline{U}(f) > u(x) > \overline{U}(g)$.
The preceding arguments imply that $f \succ x$ and $x \succ g$, from which $f \succ g$ holds by transitivity.

To prove the only if direction, suppose that $f \succ g$.
By Lemma \ref{lem_calibration}, we can find $x,y \in X$ such that $f \succ x \succ y \succ g$.
By the definitions of $\underline{U}$ and $\overline{U}$, it follows that $\underline{U}(f) \ge u(x)$ and $u(y) \ge \overline{U}(g)$.
Indeed, if $u(x) > \underline{U}(f)$, then there exists $x' \in X$ with $f \bowtie x'$ such that $u(x) > u(x')$, or $x \succ x'$.
Since $f \succ x$, in that case, transitivity dictates $f \succ x'$, a contradiction.
So, $\underline{U}(f) \ge u(x)$ holds, and similarly, one can show that $u(y) \ge \overline{U}(g)$.
Since $u(x) > u(y)$ by $x \succ y$, we conclude that $\underline{U}(f) > \overline{U}(g)$.
\end{proof}

Our next task is to show that $\underline{U}$ and $\overline{U}$ functions admit multiprior representations as in the statement.
To this end, the next series of claims translate the axioms of $\succ$ into the functional properties of $\underline{U}$ and $\overline{U}$.
In what follows, since arguments are symmetric, we only prove the properties of $\underline{U}$ (or $\underline{T}$, which will soon be defined).

\begin{claim} \label{claim_constant}
For all $f \in \calF$, $x \in X$, and $\alpha \in [0,1]$, we have $\underline{U}(\alpha f + (1-\alpha) x) = \alpha \underline{U}(f) + (1-\alpha) u(x)$ and $\overline{U}(\alpha f + (1-\alpha) x) = \alpha \overline{U}(f) + (1-\alpha) u(x)$.
\end{claim}

\begin{proof}
The claim is trivial when $\alpha \in \{0,1\}$, so let $\alpha \in (0,1)$.
Given any $\epsilon > 0$, the definition of $\underline{U}$ implies that there exists $y \in X$ with $f \bowtie y$ such that $u(y) \le \underline{U}(f) + \epsilon$.
By Axiom \ref{axiom_cindp}, $f \bowtie y$ implies $\alpha f + (1-\alpha) x \bowtie \alpha y + (1-\alpha) x$.
Again by the definition of $\underline{U}$, it follows that
\begin{align*}
\underline{U}(\alpha f + (1-\alpha) x) &\le u(\alpha y + (1-\alpha)x)
=\alpha u(y) + (1-\alpha) u(x)
\le \alpha \underline{U}(f) + (1-\alpha) u(x) + \alpha \epsilon.
\end{align*}
Since $\epsilon > 0$ is arbitrary, letting $\epsilon \to 0$ yields
\begin{align} \label{eq_claim_const}
\underline{U}(\alpha f + (1-\alpha) x) \le \alpha \underline{U}(f) + (1-\alpha) u(x).
\end{align}

For the converse direction, we first consider the case $\underline{U}(f) = \inf u(X)$.
By Claim \ref{claim_property} (\ref{claim_property2}), this means $\underline{U}(f) = u(\underline{x})$, where $\underline{x}$ is the worst outcome in $f(\Omega)$.
Then, it follows that $\alpha f(\omega) + (1-\alpha) x \succsim \alpha \underline{x} + (1-\alpha) x$ for all $\omega \in \Omega$, whence again Claim \ref{claim_property} (\ref{claim_property2}) implies that
\begin{align*}
\underline{U}(\alpha f + (1-\alpha) x) 
\ge \alpha u(\underline{x}) + (1-\alpha) u(x)
= \alpha \underline{U}(f) + (1-\alpha) u(x),
\end{align*}
as required.

Next we consider the case $\underline{U} (f) > \inf u(X)$.
Suppose not, the inequality (\ref{eq_claim_const}) is strict.
Then, since $\underline{U} (f) > \inf u(X)$, we can take some $z \in X$ with $u(z) < \underline{U}(f)$ such that
\begin{align*}
\underline{U}(\alpha f + (1-\alpha) x) < \alpha u(z) + (1-\alpha) u(x) < \alpha \underline{U}(f) + (1-\alpha) u(x).
\end{align*}
By Lemma \ref{lem_rep}, the first inequality implies $\alpha f + (1-\alpha) x \not \succ \alpha z + (1-\alpha) x$, from which Axiom \ref{axiom_cindp} leads to $f \not \succ z$.
On the other hand, by Lemma \ref{lem_rep}, $u(z) < \underline{U}(f)$ implies $f \succ z$, which contradicts to the previous conclusion.
Hence, we obtain the desired equality for the case $\underline{U} (f) > \inf u(X)$ as well.
\end{proof}

Henceforth, we normalize $u\colon X \to \R$ in such a way that $[-1,1] \subseteq u(X)$.
This is possible because $u$ is non-constant and unique up to positive affine transformations.
Hence, the set $\Xi \equiv \{u(f): f \in \calF\}$ of {\it utility acts} becomes a convex subset of $\R^\Omega$ that contains $[-1,1]^\Omega$.
Moreover, we set
\begin{alignat*}{2}
\underline{T}(\xi) = \underline{U}(f) \ \text{ and } \ \overline{T}(\xi) = \underline{U}(f) \ \text{ where } \ \xi = u(f),
\end{alignat*}
for every $\xi \in \Xi$.
Note that $\underline{T}$ and $\overline{T}$ are well-define thanks to Axiom \ref{axiom_equiv}.
Indeed, since $\xi \equiv u(f) = u(g)$ means $f(\omega) \sim g(\omega)$ for all $\omega \in \Omega$, Axiom \ref{axiom_equiv} implies $f \bowtie x$ if and only if $g \bowtie x$.
By the definitions of $\underline{U}$ and $\overline{U}$, it follows that $\underline{U}(f) = \underline{U}(g)$ and $\overline{U}(f) = \overline{U}(g)$, so $\underline{T}(\xi)$ and $\overline{T}(\xi)$ do not depend on the choice of $f$ or $g$.
Moreover, Claim \ref{claim_property} implies that $\overline{T} (\xi) \ge \underline{T} (\xi)$ for all $\xi \in \Xi$, and that $\overline{T} (r\1_\Omega) = \underline{T} (r\1_\Omega) = r$ for all $r \in u(X)$.
We next show that $\underline{T}$ and $\overline{T}$ are positively homogeneous, and hence they can be consistently extended to the whole space $\R^\Omega$.

\begin{claim} \label{claim_homo}
For all $\xi \in \Xi$ and $\lambda \ge 0$ such that $\lambda \xi \in \Xi$, we have $\underline{T}(\lambda \xi) = \lambda \underline{T}(\xi)$ and $\overline{T}(\lambda \xi) = \lambda \overline{T}(\xi)$.
Consequently, they have unique extensions to $\R^\Omega$ that preserve positive homogeneity.
\end{claim}

\begin{proof}
To avoid trivial cases, let $\lambda > 0$ and $\lambda \neq 1$.
Consider the case $\lambda \in (0,1)$.
Given $\xi \in \Xi$, let $f \in \calF$ be such that $u(f) = \xi$, and let $x_0 \in X$ be $u(x_0) = 0$.
By Claim \ref{claim_constant} and the definition of $\underline{T}$, it follows that
\begin{align*}
\underline{T}(\lambda \xi) 
= \underline{T}(u \circ (\lambda f + (1-\lambda) x_0))
= \underline{U}(\lambda f + (1-\lambda) x_0)
= \lambda \underline{U}(f) = \lambda \underline{T}(\xi).
\end{align*}
Next, consider the case $\lambda > 1$.
Since $\frac{1}{\lambda} \in (0,1)$, the previous case readily implies $\underline{T}(\xi) = \underline{T}(\frac{\lambda \xi}{\lambda}) = \frac{1}{\lambda} \underline{T}(\lambda \xi)$, from which $\underline{T} (\lambda \xi) = \lambda \underline{T}(\xi)$ is obtained.

Lastly, we claim that $\underline{T}$ can be uniquely extended to $\R^\Omega$ by preserving positive homogeneity.
Indeed, given any non-zero vector $\xi \in \R^\Omega$, we can consider the normalized vector $\tilde{\xi} = \frac{\xi}{\|\xi\|}$ so that $|\tilde{\xi} (\omega)| \le 1$.
By normalization, we have $\tilde{\xi} \in \Xi$, and hence, the value $\underline{T}(\tilde{\xi})$ is already defined.
To maintain positive homogeneity, we must have $\underline{T}(\xi) = \|\xi\|\underline{T}(\tilde{\xi})$, which provides the unique extension of $\underline{T}$ to $\R^\Omega$.
\end{proof}

Henceforth, by means of Claim \ref{claim_homo}, we assume that $\underline{T}$ and $\overline{T}$ are defined on $\R^\Omega$.
The next claim collects some other properties of $\underline{T}$ and $\overline{T}$ that we will use.

\begin{claim} \label{claim_functional}
The functions $\underline{T}, \overline{T}: \R^\Omega \to \R$ possess the following properties.
\begin{enumerate}[\rm i)]
\item For all $\xi \in \R^\Omega$ and $r,r' \in \R$, if $r \ge \xi(\omega) \ge r'$ for all $\omega$, then $\underline{T}(\xi), \overline{T}(\xi) \in [r,r']$.
\label{claim_mono}

\item For all $\xi \in \R^\Omega$, $r \in \R$, and $\lambda > 0$, $\underline{T}(\lambda \xi + r\1_\Omega) = \lambda \underline{T}(\xi) + r$ and $\overline{T}(\lambda \xi + r\1_\Omega) = \lambda \overline{T}(\xi) + r$.
In particular, $\underline{T}(r\1_\Omega) = \overline{T}(r \1_\Omega) = r$ holds for all $r \in \R$.
\label{claim_add}

\item For all $\xi,\zeta \in \R^\Omega$, $\underline{T}(\xi + \zeta) \ge \underline{T}(\xi) + \underline{T}(\zeta)$ and $\overline{T}(\xi + \zeta) \le \overline{T}(\xi) + \overline{T}(\zeta)$.
\label{claim_sub}

\item $\underline{T}$ and $\overline{T}$ are continuous with respect to $\|\cdot\|$.
\label{claim_cts}
\end{enumerate}
\end{claim}

\begin{proof}
In what follows, by Claim \ref{claim_homo}, it is without loss to assume that $\xi, \zeta \in \Xi$, so there exist $f,g \in \calF$ for which $\xi = u(f)$ and $\zeta = u(g)$.
Also, let $x,x' \in X$ be such that $u(x) = r$ and $u(x') = r'$.
First, (\ref{claim_mono}) is immediately implied from Claim \ref{claim_property} (\ref{claim_property2}).
As for (\ref{claim_add}), note that by Claim \ref{claim_constant}, we have $\underline{T}(\frac{1}{2} \xi + \frac{1}{2} r\1_\Omega) = \frac{1}{2} \underline{T}(\xi) + \frac{r}{2}$.
Multiplying both sides by $2$, it follows that $\overline{T}(\xi + r\1_\Omega) = \underline{T}(\xi) + r$ by Claim \ref{claim_homo}.
Furthermore, by this and Claim \ref{claim_homo}, we have $\overline{T}(\lambda \xi + r\1_\Omega) = \lambda \underline{T}(\xi) + r$ for any $\lambda > 0$.

As for (\ref{claim_sub}), note that $\underline{T}(\xi + \zeta) \ge \underline{T}(\xi) + \underline{T}(\zeta)$ is equivalent to $\underline{T}(\frac{1}{2}\xi + \frac{1}{2}\zeta) \ge \frac{1}{2}\underline{T}(\xi) + \frac{1}{2}\underline{T}(\zeta)$.
To show this, we first consider the case $\underline{T}(\xi) = \underline{T}(\zeta)$.
Consider any $k < \underline{T}(\xi)$, and let $k = u(y)$.
Since $\xi = u(f)$ and $\zeta = u(g)$, it follows that $f \succ y$ and $g \succ y$.
Axiom \ref{axiom_conv} then leads to $\frac{1}{2} f + \frac{1}{2} g \succ y$, and therefore, $\underline{T}(\frac{1}{2}\xi + \frac{1}{2}\zeta) = \underline{U} (\frac{1}{2} f + \frac{1}{2} g) > u(y) = k$.
Since $k$ can be arbitrarily close to $\underline{T}(\xi)$, or equivalently to $\underline{T}(\zeta)$, it follows that $\underline{T}(\frac{1}{2}\xi + \frac{1}{2}\zeta) \ge \frac{1}{2}\underline{T}(\xi) + \frac{1}{2}\underline{T}(\zeta)$.
For the general case, assume without loss of generality that $\underline{T}(\xi) > \underline{T}(\zeta)$, and denote the difference by $\delta = \underline{T}(\xi) - \underline{T}(\zeta) > 0$.
Let $\tilde{\zeta} = \zeta + \delta \1_\Omega$.
By (\ref{claim_add}), it follows that $\underline{T}(\tilde{\zeta}) = \underline{T}(\zeta) + \delta = \underline{T}(\xi)$.
Hence, by the preceding argument, we have $\underline{T}(\frac{1}{2}\xi + \frac{1}{2} \tilde{\zeta}) \ge \frac{1}{2}\underline{T}(\xi) + \frac{1}{2}\underline{T}(\tilde{\zeta})$.
Again by (\ref{claim_add}), observe that the left-side is $\underline{T}(\frac{1}{2}\xi + \frac{1}{2} \zeta) + \frac{\delta}{2}$, whereas the right-side is $\frac{1}{2}\underline{T}(\xi) + \frac{1}{2}\underline{T}(\zeta) + \frac{\delta}{2}$.
Offsetting the term $\frac{\delta}{2}$, it follows that $\underline{T}(\frac{1}{2}\xi + \frac{1}{2} \zeta) \ge \frac{1}{2}\underline{T}(\xi) + \frac{1}{2}\underline{T}(\zeta)$.

Lastly, (\ref{claim_add}) and (\ref{claim_sub}) imply $\overline{T}$ is convex and $\underline{T}$ is concave, from which (\ref{claim_cts}) obtains.
\end{proof}

We now come to combine the functional properties of $\overline{T}$ and $\underline{T}$ obtained so far to establish their integral representations.
Given any $\xi \in \R^\Omega$, we define the {\it upper} and {\it lower contour sets} of $\xi$ as follows:
\begin{alignat*}{8}
&\calU (\xi) & &= \left\{\zeta \in \R^\Omega: \underline{T} (\zeta) > \overline{T} (\xi) \right\}, \hspace{12pt}
& &\calU_{\rm diag} (\xi) & &= \calU(\xi) \cap \R^\Omega_{\rm diag}, \\
&\calL (\xi) & &= \left\{\zeta \in \R^\Omega: \underline{T} (\xi) > \overline{T} (\zeta) \right\},
& &\calL_{\rm diag}(\xi) & &= \calL(\xi) \cap \R^\Omega_{\rm diag}.
\end{alignat*}
Moreover, let $\xi \mapsto \overline{\xi}$ and $\xi \mapsto \underline{\xi}$ be the operators from $\R^\Omega$ to $\R^\Omega_{\rm diag}$ defined by
\begin{alignat*}{2}
\overline{\xi} &= \argmin & &\bigl\{\overline{T}(\zeta): \zeta \in \cl (\calU_{\rm diag} (\xi)) \bigr\}, \\
\underline{\xi} &= \argmax & &\bigl\{\underline{T}(\zeta): \zeta \in \cl (\calL_{\rm diag} (\xi)) \bigr\},
\end{alignat*}
which are well-defined since $\underline{T}(r\1_\Omega) = \overline{T}(r\1_\Omega) = r$ is one-to-one from $\R^\Omega_{\rm diag}$ to $\R$.

The next claim collects some important properties of contour sets.
In particular, it proves that (\ref{contour3}) the contour sets of any vector $\xi$ are invariant to the operators $\xi \mapsto \overline{\xi}, \underline{\xi}$, and that (\ref{contour4}) the contour sets of constant vectors can be shifted along the $45^\circ$ line.
By means of these observations, it is enough for us to characterize the contour sets of the origin, $\calU(0)$ and $\calL(0)$, to recover those of arbitrary points.

\begin{claim} \label{claim_contour}
The following are true for any $\xi, \zeta \in \R^\Omega$ and $\lambda \in \R$.
\begin{enumerate}[\rm (a)]
\item $\calU$ is decreasing in $\overline{T}$ in the sense that $\calU(\xi) \subseteq \calU(\zeta)$ whenever $\overline{T}(\xi) \ge \overline{T}(\zeta)$.
\label{contour1}

\item $\calL$ is increasing in $\underline{T}$ in the sense that $\calL(\xi) \supseteq \calL(\zeta)$ whenever $\underline{T}(\xi) \ge \underline{T}(\zeta)$.
\label{contour2}

\item $\calU(\xi) = \calU(\overline{\xi})$ and $\calL(\xi) = \calL(\underline{\xi})$.
\label{contour3}

\item $\calU(\lambda \1_{\Omega}) = \calU(0) + \lambda \1_{\Omega}$ and $\calL(\lambda \1_{\Omega}) = \calL(0) + \lambda \1_{\Omega}$.
\label{contour4}

\item $\calU(0)$ and $\calL(0)$ are open convex cones such that $\R^\Omega_{++} \subseteq \calU(0)$ and $\R^\Omega_{--} \subseteq \calL(0)$.
\label{contour5}
\end{enumerate}
\end{claim}

\begin{proof}
Note that (\ref{contour1}) and (\ref{contour2}) are obvious from the definitions.
So, in order to prove (\ref{contour3}), it is enough to show that $\overline{T}(\xi) = \overline{T} (\overline{\xi})$ and $\underline{T}(\xi) = \underline{T} (\underline{\xi})$ hold for every $\xi \in \R^\Omega$.
Clearly, the definitions of the operators $\xi \mapsto \overline{\xi}, \underline{\xi}$ imply that $\overline{T}(\xi) \le \overline{T} (\overline{\xi})$ and $\underline{T}(\xi) \ge \underline{T} (\underline{\xi})$.
Suppose, as a way of contradiction, that $\overline{T}(\xi) < \overline{T} (\overline{\xi})$ holds, so we can pick a number $r \in (\overline{T}(\xi),\overline{T} (\overline{\xi}))$.
It follows that
\begin{align*}
\overline{T}(\overline{\xi}) > \overline{T}(r\1_\Omega) = r = \underline{T}(r\1_\Omega) > \overline{T}(\xi),
\end{align*}
which means that there exists a constant vector $r\1_\Omega \in \calU_{\rm diag} (\xi)$ smaller than $\overline{\xi}$.
This is a contradiction to the minimality of $\overline{\xi}$.
Hence, $\overline{T}(\xi) = \overline{T} (\overline{\xi})$ holds.
Analogously, we can show that $\underline{T}(\xi) = \underline{T} (\underline{\xi})$.

As for (\ref{contour4}), consider any $\xi \in \R^\Omega$ and $\lambda \in \R$.
Note that Claim \ref{claim_functional} (\ref{claim_add}) implies $\underline{T} (\xi - \lambda \1_\Omega) = \underline{T} (\xi) - \lambda$.
So, it follows that
\begin{align*}
\xi \in \calU(\lambda \1_{\Omega})
\Longleftrightarrow \underline{T}(\xi) > \lambda
&\Longleftrightarrow \underline{T}(\xi - \lambda \1_{\Omega}) > 0 \\
&\Longleftrightarrow (\xi - \lambda \1_{\Omega}) \in \calU(0)
\Longleftrightarrow \xi \in \calU(0) + \lambda \1_{\Omega}.
\end{align*}
Since $\xi$ is arbitrary, we obtain $\calU(\lambda \1_{\Omega}) = \calU(0) + \lambda \1_{\Omega}$.
Similarly, one can show that $\calL(\lambda \1_{\Omega}) = \calL(0) + \lambda \1_{\Omega}$.

As for (\ref{contour5}), observe that Claim \ref{claim_functional} implies that $\calU(0)$ and $\calL(0)$ are open by (\ref{claim_cts}), conic by  (\ref{claim_add}), and convex by (\ref{claim_add}) and (\ref{claim_sub}).
Moreover, $\R^\Omega_{++} \subseteq \calU(0)$ and $\R^\Omega_{--} \subseteq \calL(0)$ hold by (\ref{claim_mono}).
\end{proof}

Now, we characterize the convex closed cones $\calU(0)$ and $\calL(0)$ as the intersections of their supporting hyperplanes.
The following result is based on standard duality arguments.

\begin{claim} \label{SHP}
There exist non-empty, closed, and convex sets $C, D \subseteq \Delta (\Omega)$  such that
\begin{align}
&\calU(0) = \biggl\{ \xi \in \R^\Omega: \min_{\mu \in C} \int \xi \d\mu > 0 \biggr\}, \label{upper0} \\
&\calL(0) = \biggl\{ \xi \in \R^\Omega: \max_{\mu \in D} \int \xi \d\mu > 0 \biggr\}. \label{lower0}
\end{align}
\end{claim}

\begin{proof}
Consider any $\zeta \in \R^\Omega \setminus \cl (\calU(0))$.
Since Claim \ref{claim_contour} (\ref{contour5}) implies $\cl (\calU(0))$ is closed and convex, the separating hyperplane theorem yields a non-zero vector $\mu_\zeta \in \R^\Omega$ such that
\begin{align*}
b \equiv \min_{\xi \in \cl(\calU(0))} \langle \mu_\zeta, \xi \rangle > \langle \mu_\zeta, \zeta \rangle.
\end{align*}
Claim \ref{claim_contour} (\ref{contour5}) also implies $\cl(\calU(0))$ is conic and contains $\R^\Omega_+$.
Thus $\mu_{\zeta} \ge 0$, and hence we can normalize it to be a probability measure.
Note that $b \le 0$ because $0 \in \cl(\calU(0))$.
Moreover, if $b < 0$, then there exists some $\xi \in \cl(\calU(0))$ such that $\langle \xi, \mu_\zeta\rangle < 0$, but then the above left-side expression could be made arbitrarily small, a contradiction.
Thus $b = 0$.

Now we set $C = \cco \bigl( \{\mu_{\zeta}: \zeta \in \R^\Omega \setminus \cl(\calU(0)) \} \bigr)$, which is a non-empty convex compact subset of $\Delta(\Omega)$ .
Note that $\min_{\mu \in C} \int \xi \d\mu \ge 0$ for every $\xi \in \cl(\calU(0))$, whereas $\min_{\mu \in C} \int \zeta \d\mu < 0$ for every $\zeta \notin \cl (\calU(0))$ since $\mu_{\zeta} \in C$.
This implies that
\begin{align} \label{upper00}
\cl(\calU(0)) = \biggl\{ \xi \in \R^\Omega: \min_{\mu \in C} \int \xi \d\mu \ge 0 \biggr\}.
\end{align}
One can easily show that the interior of the right-side (\ref{upper00}) can be obtained by replacing strict inequality with weak inequality, i.e., it equals the right-side of (\ref{upper0}).
On the other hand, the interior of $\cl (\calU(0))$ coincides with $\calU(0)$ by the fact that $\calU(0)$ is open and convex from Claim \ref{claim_contour} (\ref{contour5}).
Together these observations, we conclude that $\calU(0)$ is characterized as in (\ref{upper0}).
Analogously, we can characterize $\calL(0)$ as in (\ref{lower0}).
\end{proof}

Note that by definition, $\overline{\zeta}$ is a constant vector to which $\underline{T}$ assigns the value weakly greater than $\overline{T}(\zeta)$.
This implies that $\zeta \in \cl \calL(\overline{\zeta})$, from which Claims \ref{claim_contour} and \ref{SHP} together imply that
\begin{align}
\zeta \in \cl(\calL(\overline{\zeta}))
= \cl(\calL(0)) + \overline{\zeta} \nonumber
&\Longleftrightarrow \zeta - \overline{\zeta} \in \cl(\calL(0)) \nonumber \\
&\Longleftrightarrow \max_{\mu \in D} \int (\zeta - \overline{\zeta}) \d\mu \le 0
\Longleftrightarrow \max_{\mu \in D} \int \zeta \d\mu \le \overline{\zeta}, \label{zeta_conc}
\end{align}
where $\overline{\zeta}$ is identified with the corresponding real number.
We claim that the last inequality in (\ref{zeta_conc}) must be tight.
Indeed, if it were not, we could take some $\zeta_0 \in \R^\Omega_{\rm diag}$ such that $\max_{\mu \in D} \int \zeta d\mu < \zeta_0 < \overline{\zeta}$.
Again by Claims \ref{claim_contour} and \ref{SHP}, the following equivalence holds:
\begin{align*}
\zeta \in \calL(\zeta_0) = \calL(0) + \zeta_0
&\Longleftrightarrow \zeta - \zeta_0 \in \calL(0) \\
&\Longleftrightarrow \max_{\mu \in D} \int (\zeta - \zeta_0) \d \mu < 0
\Longleftrightarrow \max_{\mu \in D} \int \zeta \d \mu < \zeta_0.
\end{align*}
As such, since the last assertion is true by the assumption, we have $\zeta \in \calL(\zeta_0)$, or equivalently, $\zeta_0 \in \calU(\zeta)$.
However, since $\overline{\zeta} > \zeta_0$, we encounter a contradiction to the minimality of $\overline{\zeta}$.
Therefore, we have shown that every $\zeta \in \R^\Omega$ satisfies the following identity:\footnote{
Though we do not use it in the rest of the proof, a symmetric identity $\underline{\zeta} = \min_{\mu \in C} \int \zeta \d\mu$ holds as well.}
\begin{align}
\overline{\zeta} = \max_{\mu \in D} \int \zeta \d\mu. \label{conc1}
\end{align}

Now, fix any $f, g \in \calF$, and let $\xi = u(f)$ and $\zeta = u(g)$.
By Lemma \ref{lem_rep} and the definitions of $\underline{T}$ and $\overline{T}$, it follows that $f \succ g$ if and only if $\underline{T}(\xi) > \overline{T}(\zeta)$.
Combining this with Claims \ref{claim_contour} and \ref{SHP}, we observe
\begin{align}
f \succ g &\Longleftrightarrow \underline{U}(f) > \overline{U}(g) \nonumber \\[5pt]
&\Longleftrightarrow \underline{T}(\xi) \hspace{2pt} > \overline{T}(\zeta) \nonumber \\[5pt]
&\Longleftrightarrow \xi \in \calU(\zeta) = \calU(\overline{\zeta}) = \calU(0) + \overline{\zeta} \nonumber \\[5pt]
&\Longleftrightarrow \xi - \overline{\zeta} \in \calU(0) \nonumber \\[3pt]
&\Longleftrightarrow \min_{\mu \in C} \int (\xi - \overline{\zeta}) \d\mu > 0 \nonumber \\
&\Longleftrightarrow \min_{\mu \in C} \int \xi \d\mu > \overline{\zeta}, \label{conc3}
\end{align}
where the last equivalence follows from the fact that the integral of the constant vector $\overline{\zeta}$ does not depend on the choice of $\mu \in C$.
Substituting (\ref{conc1}) into (\ref{conc3}), we obtain
\begin{align}
f \succ g \Longleftrightarrow \min_{\mu \in C} \int u(f) \d\mu > \max_{\mu \in D} \int u(g) \d\mu. \label{conclusion}
\end{align}
Therefore, we obtained the desired representation.

Our remaining task is to show that $C$ and $D$ are not disjoint.
Suppose, arguing by contradiction, that $C \cap D = \emptyset$ holds.
Then, since $C$ and $D$ are convex compact subsets of $\R^\Omega$, there exists a non-zero vector $\xi \in \R^\Omega$ such that $\min_{\mu \in C} \langle \xi, \mu \rangle > \max_{\mu \in D} \langle \xi, \mu \rangle$.
Hence, after normalizing $\xi$, we can find $f \in \calF$ and $x \in X$ such that $\min_{\mu \in C} \int u(f) \d\mu > u(x) > \max_{\mu \in D} \int u(f) \d\mu$.
It follows that $f \succ x$ and $x \succ f$, which contradicts to the asymmetry of $\succ$.
Hence, $C \cap D \neq \emptyset$.
\hfill {\it Q.E.D.}


\subsubsection{Necessity}

Let $\succ$ be a preference that admits the representation of Theorem \ref{thm_main}.
Next we show it must satisfy Axioms \ref{axiom_str}--\ref{axiom_mono}.
The restriction of $\succ$ to $X$ is non-trivial and negatively transitive since it is represented by a non-constant function $u$.
That $\succ$ is asymmetry follows from the assumption that $C \cap D \neq \emptyset$.
To prove the transitivity of $\succ$, consider acts $f \succ g \succ h$.
We have
 \begin{align*}
\min_{\mu \in C} \int u(f) \d \mu > \max_{\mu \in D} \int u(g) \d \mu
&\geq \max_{\mu \in C \cap D} \int u(g) \d \mu \\
&\geq \min_{\mu \in C \cap D} \int u(g)  \d \mu
\geq  \min_{\mu \in C}\int u(g)  \d \mu >  \max_{\mu \in D} \int (u \circ h) \d \mu,
 \end{align*}
from which $f \succ h$.
Hence, $\succ$ satisfies Axiom \ref{axiom_str}.

That $\succ$ must satisfy Axioms \ref{axiom_cts}--\ref{axiom_conv} is easy to see, as it follows immediately from the representation.
To show that $\succ$ satisfies Axiom \ref{axiom_contagion}, consider its contrapositive statement that posits:\ {\it whenever two acts $f$ and $g$ are comparable, every constant act should be comparable with at least one of them.}
Assuming $f \succ g$ without loss, we have
\begin{align*}
\min_{\mu \in C} \int u(f) \d \mu > \max_{\mu \in D} \int u(g) \d \mu.
\end{align*}
This implies for every $x \in X$, at least one of the following holds:
\begin{align*}
u(x) < \min_{\mu \in C} \int u(f) \d \mu \ \text{ or } \ u(x) > \max_{\mu \in D} \int u(g) \d \mu
\end{align*}
Since the former implies $f \succ x$ and the latter implies $x \succ g$, we obtain Axiom~\ref{axiom_contagion}.
Finally, the representation easily implies that $\succ$ satisfies the monotonicity conditions, i.e., Axioms \ref{axiom_equiv} and \ref{axiom_mono}.
\hfill {\it Q.E.D.}


\subsubsection{Uniqueness}

 We now show the sets $C$ and $D$ in the representation are unique. To this end, suppose $C_0$ and $D_0$ are two closed convex sets that satisfy the representation.
Suppose $C \neq C_0$. Without loss of generality, assume there exists $\mu^* \in C \setminus C_0$.
Since $C_0$ is convex and closed, there exists a non-zero vector $\xi \in \R^\Omega$ and $k \in \mathbb{R}$ such that
\begin{align*}
\min_{\mu \in C_0} \langle \mu, \xi \rangle > k > \langle \mu^*, \xi \rangle.
\end{align*}
By scaling $\xi$ and $k$ appropriately, and using the assumption that $[-1,1] \subseteq u(X)$, we can find $f \in \calF$ and $x \in X$ such that $u(f) = \xi$ and $u(x) = k$. 
Note that $f \succ x$ since $\min_{\mu \in C_0} \int u(f) \d \mu > u(x)$.
Yet, $\min_{\mu \in C}\int u(f) \d \mu < u(x)$ since $\mu^* \in C$, so $f \not \succ x$, a contradiction.
Thus $C$ is unique, and an analogous argument shows $D$ is unique.
\hfill {\it Q.E.D.}


\subsection{Other Proofs}

\subsubsection{Proof of Proposition \ref{prop_sym}}

Let $\succ$ be represented by a profile $(u,C,D)$, and assume that $[-1,1] \subseteq u(X)$ without loss of generality.
Since arguments are symmetric, we only prove (\ref{prop_sym1}).

Suppose that $D \subseteq C$, and we shall show that $\succ$ satisfies Axiom \ref{sym1}.
Consider any complementary acts $f,g \in \calF$ such that $\frac{1}{2}f (\omega) + \frac{1}{2}g(\omega) \sim x$ for all $\omega \in \Omega$, and assume that $f \succ x$.
Letting $\xi = u(f)$ and $\zeta = u(g)$, this implies that $\frac{1}{2}\xi + \frac{1}{2}\zeta = k\1_\Omega$ with $u(x) = k$.
Since $f \succ x$, it follows that
\begin{align*}
\min_{\mu \in C} \langle \xi, \mu \rangle > k
&\Longleftrightarrow \min_{\mu \in C} \langle \xi - k\1_\Omega, \mu \rangle > 0 \\
&\Longleftrightarrow \frac{1}{2} \min_{\mu \in C} \langle \xi - \zeta, \mu \rangle > 0
\Longleftrightarrow \max_{\mu \in C} \langle \zeta - \xi, \mu \rangle < 0,
\end{align*}
where the second equivalence follows from $\frac{1}{2}\xi + \frac{1}{2}\zeta = k\1_\Omega$, and the last is obtained by multiplying both sides by $-1$.
Moreover, since $D \subseteq C$, it follows that
\begin{align*}
\max_{\mu \in D} \langle \zeta - \xi, \mu \rangle < 0.
\end{align*}
Plugging $\xi = 2k\1_\Omega - \zeta$ into the above, we obtain
\begin{align*}
2 \max_{\mu \in D} \langle \zeta - k\1_\Omega, \mu \rangle < 0
\Longleftrightarrow \max_{\mu \in D} \langle \zeta, \mu \rangle > k.
\end{align*}
Since $\zeta = u(g)$ and $k = u(x)$, this implies $x \succ g$.
Therefore, $\succ$ satisfies Axiom \ref{sym1}.

For the converse direction, suppose that $D \not \subseteq C$, namely, there exists some $\mu^* \in D$ such that $\mu^* \notin C$.
Since $C$ is closed and convex, there exist a non-zero vector $\xi \in \R^\Omega$ and $k \in \mathbb{R}$ such that
\begin{align} \label{eq_sym}
\min_{\mu \in C} \langle \mu, \xi \rangle > k > \langle \mu^*, \xi \rangle.
\end{align}
By scaling $\xi$ and $k$ appropriately, and using the assumption that $[-1,1] \subseteq u(X)$, we can find $f,h \in \calF$ and $x \in X$ such that $u(f) = \frac{1}{2}\xi$, $u \circ h = -\xi$ and $u(y) = 2k$.
Moreover, let $g = \frac{1}{2}h + \frac{1}{2}y$.
It follows that
\begin{align*}
\frac{1}{2}f + \frac{1}{2}g = \frac{1}{4} \xi + \frac{1}{2} \left( -\frac{1}{2}\xi + k\1_\Omega \right) = \frac{k}{2}\1_\Omega,
\end{align*}
meaning that $f$ and $g$ are complementary, and that $\frac{1}{2}f(\omega)+\frac{1}{2}g(\omega) \sim x$ holds for some outcome $x \in X$ such that $u(x) = \frac{k}{2}$.
Since $u(f) = \frac{1}{2} \xi$, the equation (\ref{eq_sym}) implies that
\begin{align*}
\min_{\mu \in C} \int u(f) \d\mu
= \frac{1}{2} \min_{\mu \in C} \langle \mu, \xi \rangle
> \frac{k}{2} = u(x),
\end{align*}
from which $f \succ x$.
On the other hand, observe that $u(g) = u(\frac{1}{2} h + \frac{1}{2} y) = - \frac{1}{2} \xi + k\1_\Omega$.
Since $\mu^* \in D$, the equation (\ref{eq_sym}) implies that
\begin{align*}
\max_{\mu \in D} \int u(g) \d \mu
\ge \int u(g) \d \mu^*
= -\frac{1}{2} \langle \mu^*, \xi \rangle + k
> - \frac{k}{2} + k = \frac{k}{2} = u(x),
\end{align*}
from which $x \not \succ g$.
To sum up, we have shown that $\frac{1}{2} f (\omega)+\frac{1}{2} g(\omega) \sim x$ for all $\omega \in \Omega$, $f \succ x$, and $x \not \succ g$, which mean the violation of Axiom \ref{sym1}.
\hfill {\it Q.E.D.}


\subsubsection{Proof of Proposition \ref{prop_attitude}}

Let $\succ_1$ and $\succ_2$ be represented by a profile $(u,C_1,D_1)$ and $(u,C_2,D_2)$, respectively.
Assume that $[-1,1] \subseteq u(X)$ without loss of generality.
We only prove \eqref{prop_attitude1}. The argument for \eqref{prop_attitude2} is entirely symmetric.

Suppose that $C_2 \subseteq C_1$, and let $f \in \calF$ and $x \in X$ be such that $f \succ_1 x$.
The inclusion implies that
\begin{align*}
\min_{\mu \in C_2} \int u(f) \d \mu \ge \min_{\mu \in C_1} \int u(f) \d \mu > u(x),
\end{align*}
from which $f \succ_2 x$.
Hence, $\succ_1$ is more ambiguity averse over the alternative than $\succ_2$.

For the converse direction, suppose that $C_2 \not \subseteq C_1$, namely, there exists some $\mu^* \in C_2$ such that $\mu^* \notin C_1$.
Since $C_1$ is closed and convex, there exist a non-zero vector $\xi \in \R^\Omega$ and $k \in \mathbb{R}$ such that
\begin{align} \label{eq_attitude}
\min_{\mu \in C_1} \langle \mu, \xi \rangle > k > \langle \mu^*, \xi \rangle.
\end{align}
By scaling $\xi$ and $k$ appropriately, and using the assumption that $[-1,1] \subseteq u(X)$, we can find $f \in \calF$ and $x \in X$ such that $u(f) = \xi$ and $u(x) = k$.
Thus \eqref{eq_attitude} implies $f \succ_1 x$.
On the other hand, since $\mu^* \in C_2$, \eqref{eq_attitude} also implies $f \not \succ_2 x$.
This shows that $\succ_1$ is {\it not} more ambiguity averse over the alternative than $\succ_2$.
\hfill {\it Q.E.D.}


\subsubsection{Proof of Corollary \ref{cor_attitude}}

Let $\succ_1$ and $\succ_2$ be represented by a profile $(u,C_1,D_1)$ and $(u,C_2,D_2)$, respectively.
If $\succ_1$ is more conservative than $\succ_2$, then by definition, $\succ_1$ must be both more ambiguity averse over the alternative and more ambiguity loving over the status-quo than $\succ_2$.
Then Proposition \ref{prop_attitude} readily implies $C_2 \subseteq C_1$ and $D_2 \subseteq D_1$.

Conversely, assume $C_2 \subseteq C_1$ and $D_2 \subseteq D_1$ hold.
Again by Proposition \ref{prop_attitude}, this implies $\succ_1$ is both more ambiguity averse over the alternative and more ambiguity loving over the status quo than $\succ_2$.
Consider any acts $f,g \in \calF$ such that $f \succ_1 g$.
Since $\succ_1$ satisfies Axiom \ref{axiom_contagion}, there exists some $x \in X$ for which $f \succ_1 x \succ_1 g$.
It follows that $f \succ_2 x \succ_2 g$ because $\succ_1$ is both more ambiguity averse and loving than $\succ_2$, and thus, we have $f \succ_2 g$ by the transitivity of $\succ_2$.
Therefore, $\succ_1$ is more conservative than $\succ_2$.
\hfill {\it Q.E.D.}


\subsubsection{Proof of Proposition \ref{thm:connections}}

Consider a twofold multiprior preference $\succ$ with representation $(u, C, D)$.
As usual, assume that $[-1,1] \subseteq u(X)$.
We want to show that the following are equivalent:
\begin{enumerate}[(a)]
\item \label{it:conn-a}
$\succ$ satisfies monotonicity, i.e., Axiom \ref{ax:standard-monotonicity}.
\item \label{it:conn-b}
$\succ$ satisfies independence, i.e., Axiom \ref{ax:standard-independence}.
\item \label{it:conn-c}
$C=D=\{\mu\}$ for some $\mu \in \Delta (\Omega)$.
\end{enumerate}

Clearly, (\ref{it:conn-c}) implies both (\ref{it:conn-a}) and (\ref{it:conn-b}).
Let us show that (\ref{it:conn-b}) implies (\ref{it:conn-a}).
Consider any acts $f,g \in \calF$ such that $f(\omega) \succ g(\omega)$ for all $\omega \in \Omega$.
We can let $u(f), u(g) \in [-1,1]^\Omega$ without loss of generality.
Let $u(x_0) = 0$, and let $\hat{g} \in \calF$ be an act such that $u(\hat{g}) = - u(g)$.
So, $\frac{1}{2}g(\omega) + \frac{1}{2}\hat{g}(\omega) \sim x_0$ holds for all $\omega \in \Omega$.
Moreover, since $u$ is affine, we have $\frac{1}{2}u(f(\omega))+\frac{1}{2}u(\hat{g}(\omega)) > u(x_0)$ for all $\omega \in \Omega$, from which Axiom \ref{axiom_mono} implies $\frac{1}{2}f+\frac{1}{2}\hat{g} \succ x_0$.
By Axiom~\ref{axiom_equiv}, it follows that $\frac{1}{2}f+\frac{1}{2}\hat{g} \succ \frac{1}{2}g+\frac{1}{2}\hat{g}$.
Thus, Axiom \ref{ax:standard-independence} implies $f \succ g$, as desired.

To complete the proof of (\ref{it:conn1}), let us show that (\ref{it:conn-a}) implies (\ref{it:conn-c}).
Let $f$ be any act such that $u(f) \in (-1,1)^\Omega$.
Given any small $\epsilon > 0$, let $f_\epsilon \in \calF$ be an act such that $u(f_\epsilon) = u(f) - \epsilon$.
Axiom \ref{ax:standard-monotonicity} implies $f \succ f_\epsilon$, i.e., $\min_{\mu \in C} \int u(f) \d p > \max_{p \in D} \int u(f) \d p - \epsilon$.
Letting $\epsilon \to 0$, we obtain the inequality
\begin{align*}
\min_{\mu \in C} \int u(f) \d\mu \ge \max_{\mu \in D} \int u(f) \d\mu,
\end{align*}
while we also have
\begin{align*}
\max_{\mu \in D} \int u(f) \d\mu \ge \min_{\mu \in C} \int u(f) \d\mu,
\end{align*}
since $C \cap D \neq \emptyset$.
Therefore, we have shown that $\min_{\mu \in C} \int \xi \d\mu = \max_{\mu \in D} \int \xi \d\mu$ for every vector $\xi \in (-1,1)^\Omega$.
It now follows from a standard separation argument that this is possible only if there exists $\mu \in \Delta(\Omega)$ such that $C = D = \{\mu\}$.

Now we turn to statement~(\ref{it:conn2}).
Let $\succ$ be a Bewley preference, with a representation $(u, C)$, satisfying Axiom~\ref{axiom_contagion}.
It is immediate to verify that $\succ$ satisfies all the remaining axioms in Theorem~\ref{thm_main}, meaning that $\succ$ is a twofold multiprior preference.
But being $\succ$ a Bewley preference, it also satisfies the independence and the monotonicity axioms.
Thus, by statement~(\ref{it:conn1}), it must be a subjective expected utility preference.
\hfill {\it Q.E.D.}


\subsubsection{Proof of Proposition \ref{thm:Bewley-Interval-Comparison}}

Let $\succ$ be a twofold conservative preference with representation $(u, C, D)$, and let $\succ^*$ be a Bewley preference $\succ^*$ with representation $(u, C^*)$.
First, suppose that $C^* \subseteq C \cap D$. 
If $f \succ g$, then
\begin{align*}
\min_{\mu \in C^*} \int u(f) \d\mu
\ge \min_{\mu \in C} \int u(f) \d\mu
> \max_{\mu \in D} \int u(g) \d\mu
\ge \max_{\mu \in C^*} \int u(g) \d\mu,
\end{align*}
from which $f \succ^* g$.
Hence, $\succ$ is more conservative than $\succ^*$.

Conversely, suppose $\succ$ is more conservative than $\succ^*$.
We want to show that $C^* \subseteq C \cap D$.
Suppose, arguing by contradiction, that there exists $\mu^* \in C^*\setminus C$.
By standard separation arguments, we can find an act $f \in \calF$ and $k \in [-1,1]$ such that 
\begin{align*}
\min_{\mu \in C} \int u(f) \d \mu > k > \int u(f) \d \mu^*.
\end{align*}
Letting $x \in X$ be such that $u(x)=k$, it follows that $f \succ x$ but $f \not\succ^* x$, a contradiction. 
By the same logic, suppose there exists $\mu^* \in C^* \setminus D$. We can find an act $f \in \calF$ and $x \in X$ such that
\begin{align*}
\int u(f) \d \mu^* > u(x) > \max_{\mu \in D} \int u(f) \d \mu.
\end{align*}
In this case, we have $x \succ f$ but $x \not\succ^* f$, which leads to another contradiction.
Therefore, we must have $C^* \subseteq C \cap D$. 
\hfill {\it Q.E.D.}


\subsubsection{Proof of Proposition \ref{prop_completion}}

Consider a twofold multiprior preference $\succ$ represented by a profile $(u,C,D)$, and let $\hat{\succ}$ be an asymmetric binary relation.
First, suppose $\hat{\succ}$ is represented by a utility function $I\colon \calF \to \R$ as in (\ref{eq_alpha}) for some function $\alpha\colon \calF \to [0,1]$.
Clearly, $\hat{\succ}$ is negatively transitive.
Note that $I(x) = u(x)$ for all $x \in X$, and hence, $\hat{\succ}$ and $ \succ$ agree on $X$.
Moreover, when $f \succ g$, it holds that
\begin{align*}
I(f) \ge \min_{\mu \in C} \int u(f) \d\mu > \max_{\mu \in D} \int u(g) \d\mu \ge I(g),
\end{align*}
where the first and third inequalities follow from the fact that $C \cap D \neq \emptyset$, and that $\alpha$ takes values in $[0,1]$.
Thus $f \hat{\succ} g$ holds.
This shows that $\hat{\succ}$ is a completion of $\succ$.
Lastly, one can easily show that $\hat{\succ}$ satisfies (R1) and (R2) from the fact that $I(x) = u(x)$ for all $x \in X$.

Conversely, suppose that $\hat{\succ}$ is a regular completion of $\succ$.
Define $\hat{\succeq}$ by $f \hat{\succeq} g$ if and only if $g \not \hat{\succ} f$, and denote by $\hat{\sim}$ the symmetric part of $\hat{\succeq}$.
Since $\hat{\succ}$ is asymmetric and negatively transitive, $\hat{\succeq}$ is complete and transitive.
As is standard, we can show that every act $f$ admits a $\hat{\succeq}$-certainty equivalent.

\begin{claim} \label{claim_completion}
For any $f \in \calF$, there exists $x_f \in X$ such that $f \hat{\sim} x_f$.
Moreover, if $f \hat{\sim} x$ and $f \hat{\sim} x'$, then $u(x) = u(x')$.
\end{claim}

\begin{proof}
Since $f(\Omega)$ is finite, we can find $\overline{x}, \underline{x} \in f(\Omega)$ such that $\overline{x}\ \hat\succeq\ f(\omega)\ \hat\succeq\ \underline{x}$ for all $\omega \in \Omega$.
By (R2), it follows that $\overline{x}\ \hat\succeq\ f\ \hat\succeq\ \underline{x}$.
Moreover, by (R1) and the definition of $\hat\succeq$, the sets $\Lambda = \{\lambda \in [0,1]: \lambda \overline{x} + (1-\lambda) \underline{x}\ \hat{\succeq}\ f\}$ and $\Lambda' = \{\lambda \in [0,1]: f\ \hat{\succeq}\ \lambda \overline{x} + (1-\lambda) \underline{x}\}$ are closed relative to $[0,1]$.
Since $\Lambda \cup \Lambda' = [0,1]$ by the completeness of $\hat\succeq$, it follows that $\Lambda \cup \Lambda' \neq \emptyset$.
Hence, $x_f\ \hat\sim\ f$ holds if we set $x_f = \lambda \overline{x} + (1-\lambda) \underline{x}$ for $\lambda \in \Lambda \cup \Lambda'$.
Clearly, $u(x_f)$ does not depend on the choice of certainty equivalents, since otherwise, $\hat\sim$ would violate transitivity.
\end{proof}

Now we set $I(f) = u(x_f)$ for all $f \in \calF$.
Claim \ref{claim_completion} implies that $I$ is well-defined and represents $\succ$.
It remains to show that $I$ takes the form as in (\ref{eq_alpha}).
As such, since $\alpha(f)$ can take any value in $[0,1]$, and since $I(f) = u(x_f)$, it suffices to show that 
\begin{align} \label{eq_pf_alpha}
\min_{\mu \in C} \int u(f) \d \mu \le u(x_f) \le \max_{\mu \in D} \int u(f) \d \mu
\end{align}
holds for all $f \in \calF$.
Indeed, $u(x_f) > \min_{\mu \in C} \int u(f) \d \mu$ leads to $x_f \succ f$, but then, we have $x_f\ \hat{\succ}\ f$ since $\succ \ \subseteq \ \hat{\succ}$.
This is a contradiction to $x_f\ \hat\sim\ f$.
Similarly, $u(x_f) > \max_{\mu \in D} \int u(f) \d \mu$ leads to $x_f \succ f$, which in turn implies $x_f\ \hat\succ\ f$, a contradiction to $x_f\ \hat\sim\ f$.
Therefore, the equation (\ref{eq_pf_alpha}) holds for all $f \in \calF$, as desired.
\hfill {\it Q.E.D.}


\subsubsection{Proof of Corollary \ref{cor_maximin}}

We prove that if $(\succ,\hat{\succ})$ jointly satisfies caution, then it admits the maxmin representation $(u,C)$. The converse implication is immediate to verify.
In view of Proposition~\ref{prop_completion}, we want to show that caution implies $\alpha(f)=0$ for all $f \in \calF$.
Clearly, we can set $\alpha$ in that way for any $f$ whose expected utility does not depend on $\mu \in C$.
Now, suppose by contradiction that there exists some $f$ such that $\alpha (f) > 0$ and $\min_{\mu \in C} \int u(f) \d\mu < \max_{\mu \in C} \int u(f) \d\mu$.
Note that for every $x \in X$, if $\min_{\mu \in C} \int u(f) \d\mu < u(x) < \max_{\mu \in C} \int u(f) \d\mu$, we have $f \not \succ x$.
On the other hand, we can choose $x$ so that $u(x)$ is arbitrarily close to $\min_{\mu \in C} \int u(f) \d\mu$.
Since $\alpha > 0$, we eventually have $f \succ^* x$, from which we encounter a contradiction to caution.
Hence $\alpha (f) = 1$ holds for all $f \in \calF$, as desired.
The case of maximax is similarly discussed.
\hfill {\it Q.E.D.}

\subsubsection{Proof of Proposition \ref{thm:choice}}

Fix a constant status quo $x \in X$, and assume that a Bewley preference $\succ_{\rm B}$ with representation $(u, C)$ weakly rationalizes a choice function $c$ under $x$.
We want to show that a twofold multiprior preference $\succ_{\rm TF}$ with representation $(u,C,C)$ weakly rationalizes $c$ under $x$ as well.
To this end, fix any menu $A \in \calA$, and consider the following two cases:

\begin{itemize}
\item If $c(A;x) = x$, we have $f \not \succ_{\rm B} x$ for all $f \in A$ since $\succ_{\rm B}$ weakly rationalizes $c$ under $x$.
Note that Proposition~\ref{thm:Bewley-Interval-Comparison} implies $\succ_{\rm TF}$ is a subrelation of $\succ_{\rm B}$, so it follows that $f \not \succ_{\rm TF} x$ for all $f \in A$.

\item If $c(A;x) = f \neq x$, we have $f \succ_{\rm B} x$ and $g \not \succ_{\rm B} f$ for all $g \in A$.
Again, by Proposition~\ref{thm:Bewley-Interval-Comparison}, it follows that $g \not \succ_{\rm TF} f$ for all $g \in A$.
Moreover, notice that $f \succ_{\rm B} x$ means $\int u(f) \d \mu > u(x)$ for all $\mu \in C$, which is equivalent to that $\min_{\mu \in C} \int u(f) \d \mu > u(x)$.
This implies $f \succ_{\rm TF} x$.
\end{itemize}

\noindent
By the definition of weak rationalizability, the above discussion concludes that $\succ_{\rm TF}$ weakly rationalizes $c$ under $x$ as well.

We now prove the existence of a choice function $c$ weakly rationalizable by a twofold preference, but not by any Bewley preference.
Given any $u$ fixed, assume that $u(x) < \sup u(X)$.
Consider a twofold preference $\succ_{\rm TF}$ with representation $(u, C, C)$, in which  $C$ is not a singleton.
Then, let $f$ be an act such that $u(f)$ is contained in the open rectangular $(u(x),\, \sup u(X))^\Omega$, and such that $\int u(f) \d \mu$ is not constant over $C$.
Clearly, there exists an $\epsilon$ such that one can define the act $g$ such that $u(g) = u(f) + \epsilon \1_\Omega$, and such that $g \not\succ_{\rm TF} f$.
Now defined a choice function $c$ such that $c(\{f,g\}; x) = f$ and $c$ is weakly rationalizable by $\succ_{\rm TF}$.
This is clearly possible since $f \succ_{\rm TF} x$ and $g \not\succ_{\rm TF} f$.
But $c$ is not weakly rationalizable by any Bewley preference $\succ_{\rm B}$ since monotonicity dictates $g \succ_{\rm B} f$.
\hfill {\it Q.E.D.}

\subsubsection{Proof of Propositions \ref{prop_auc1} and \ref{obvious}}
Recall that $U(\cdot,\mu)$ is maximized at $v$ for any $\mu$, and so is $\max_{\mu \in D} U(\cdot,\mu)$.
This implies $v$ is not dominated by any $b \in V$, meaning that $v \in B^*(C, D)$.
Moreover, note that $\min_{\mu \in C} U(\cdot,\mu)$ is maximized at $v$, so any bid $b \neq v$ is dominated by some other $b'$ if and only if $b$ is dominated by $v$.
This implies we have $b \in B^*(C, D)$ if and only if $\overline{U}(b) \geq \underline{U}(v)$, where
\begin{align*}
\overline{U}(b) \equiv \max_{\mu \in D} \underbrace{\int_0^{b} (v-\omega) \d\mu(\omega)}_{\equiv U(b,\, \mu)}
\ \text{ and } \
\underline{U}(v) \equiv \min_{\mu \in C} \int_0^{v} (v-\omega) \d\mu(\omega).
\end{align*}
Given any $\mu$, note that $U(b,\mu)$ is weakly increasing on $\{0,\ldots,v\}$ and weakly decreasing on $\{0,\ldots,v\}$.
It follows that $\overline{U}(b)$, defined as the upper envelop of $U(b,\mu)$, is also weakly increasing on $\{0,\ldots,v\}$ and weakly decreasing on $\{0,\ldots,v\}$.
This implies the set of $b$ such that $\overline{U}(b) \ge \underline{U}(v)$ takes the form of a discrete interval as in Proposition \ref{prop_auc1}.

Now, assume that $C = D = \Delta (\Omega)$, and consider belief $\bar \mu$ that assigns a mass to a point $\bar b \in \Omega$.
According to $\bar \mu$, truth-telling achieves the minimal expected utility of $U(v,\bar \mu)=0$.
On the other hand, the maximal expected utility of every bid $b$ is no less than $0$ since $D = \Delta (\Omega)$.
This implies every $b$ is undominated.
\hfill {\it Q.E.D.}

\subsubsection{Proof of Theorem \ref{thm_maxmin}}

That (ii) implies (i) can be easily checked, and hence we omit the proof.
Conversely, assume that $\succ$ satisfies Axioms~\ref{axiom_str}--\ref{axiom_cindp}, \ref{axiom_contagion}--\ref{axiom_mono}, and  \ref{axiom_averse}.
Note that in the sufficiency proof of Theorem \ref{thm_main}, Axiom~\ref{axiom_conv} is not used until we prove Claim \ref{claim_functional}.
Hence, even when Axiom~\ref{axiom_conv} is replaced by Axiom~\ref{axiom_averse}, we still have a non-constant affine function $u: X \to \R$, and positively homogeneous functions $\underline{T}, \overline{T}: \R^\Omega \to \R$ with $\overline{T} \ge \underline{T}$ such that for all $f,g \in \calF$,
\begin{align} \label{eq_pf_maxmin}
f \succ g \Longleftrightarrow \underline{T}(u(f)) > \overline{T}(u(g)).
\end{align}
We can also establish several properties of $\underline{T}$ and $\overline{T}$ by mimicking the proof of Claim \ref{claim_functional}.
Specifically, the third brick point in the next claim is the only difference from before, i.e., Axiom~\ref{axiom_averse} now makes both $\underline{T}$ and $\overline{T}$ superadditive.\footnote{Evidently, if one adopts Axiom~\ref{axiom_loving} in the place of Axiom~\ref{axiom_averse}, the subadditivity of $\underline{T}$ and $\overline{T}$  will be instead obtained.
As a result, in that case, we will have max-operators in the equation (\ref{contour_maxmin2}). Together with the subsequent proof of Theorem \ref{thm_maxmin}, this illustrates the sketch of the proof of Theorem \ref{thm_maxmax}.}

\begin{claim} \label{claim_functional2}
The functions $\underline{T}, \overline{T}: \R^\Omega \to \R$ possess the following properties.
\begin{enumerate}[\rm i)]
\item For all $\xi \in \R^\Omega$ and $r,r' \in \R$, if $r \ge \xi(\omega) \ge r'$ for all $\omega \in \Omega$, then $\underline{T}(\xi), \overline{T}(\xi) \in [r,r']$.
\label{claim_mono2}

\item For all $\xi \in \R^\Omega$, $r \in \R$, and $\lambda > 0$, $\underline{T}(\lambda \xi + r\1_\Omega) = \lambda \underline{T}(\xi) + r$ and $\overline{T}(\lambda \xi + r\1_\Omega) = \lambda \overline{T}(\xi) + r$.
In particular, $\underline{T}(r\1_\Omega) = \overline{T}(r \1_\Omega) = r$ holds for all $r \in \R$.
\label{claim_add2}

\item For all $\xi,\zeta \in \R^\Omega$, $\underline{T}(\xi + \zeta) \ge \underline{T}(\xi) + \underline{T}(\zeta)$ and $\overline{T}(\xi + \zeta) \ge \overline{T}(\xi) + \overline{T}(\zeta)$.
\label{claim_sub2}

\item $\underline{T}$ and $\overline{T}$ are continuous with respect to the sup norm $\|\cdot\|$ on $\R^\Omega$.
\label{claim_cts2}
\end{enumerate}
\end{claim}

\begin{proof}
Here, we only verify the superadditivity of $\overline{T}$ since the proofs of other properties are the same as Claim \ref{claim_functional}.
As before, it is enough to show that $\overline{T}(\frac{1}{2} \xi + \frac{1}{2} \zeta) \ge \frac{1}{2} \overline{T}(\xi) + \frac{1}{2} \overline{T}(\zeta)$ holds whenever $\overline{T}(\xi) = \overline{T}(\zeta)$, because $\overline{T}$ is positively homogeneous and satisfies (\ref{claim_add2}).
In addition, by positive homogeneity, we can assume without loss that $\xi, \zeta \in (u(X))^\Omega$, so there exist $f,g \in \calF$ such that $\xi = u(f)$ and $\zeta = u(g)$.
Let $x \in X$ be such that $\overline{T}(\xi) = u(x)$.
By (\ref{eq_pf_maxmin}), we have $x \not \succ f$ and $x \not \succ g$, from which Axiom~\ref{axiom_averse} implies $x \not \succ \frac{1}{2} f + \frac{1}{2} g$.
Again by (\ref{eq_pf_maxmin}), it follows that $\overline{T}(\frac{1}{2} \xi + \frac{1}{2} \zeta) \ge u(x)$, while the right-side equals to $u(x) = \frac{1}{2} \overline{T}(\xi)+ \frac{1}{2} \overline{T}(\zeta)$.
\end{proof}

Now we define
\begin{align} \label{contour_maxmin1}
\calU (0) = \left\{\xi \in \R^\Omega: \underline{T} (\xi) \ge 0 \right\} \ \text{ and } \ 
\calV (0) = \left\{\xi \in \R^\Omega: \overline{T} (\xi) \ge 0 \right\}.
\end{align}
The properties of $\underline{T}$ and $\overline{T}$ are translated to the properties of $\calU(0)$ and $\calV(0)$ as follows.
That $\calU(0) \subseteq \calV(0)$ follows from $\overline{T} \ge \underline{T}$.
In addition, Claim \ref{claim_functional2} implies $\calU(0)$ and $\calV(0)$ are closed by (\ref{claim_cts2}), conic by (\ref{claim_add2}), convex by (\ref{claim_add2}) and (\ref{claim_sub2}), and both contain $\R^\Omega_{++}$ by (\ref{claim_mono2}).
Thus, by arguments similar to Claim \ref{SHP}, we obtain non-empty closed convex sets $C,D \subseteq \Delta(\Omega)$ such that
\begin{align} \label{contour_maxmin2}
\calU(0) = \left\{\xi \in \R^\Omega: \min_{\mu \in C} \int \xi \d \mu \ge 0  \right\} \ \text{ and } \ 
\calV(0) = \left\{\xi \in \R^\Omega: \min_{\mu \in D} \int \xi \d \mu \ge 0  \right\}.
\end{align}
Furthermore, $\calU(0) \subseteq \calV(0)$ implies $D \subseteq C$.
Indeed, if there exists $\mu^* \in D$ such that $\mu^* \notin C$, then there exists a non-zero vector $\xi \in \R^\Omega$ such that
\begin{align*}
\min_{\mu \in C} \int \xi \d \mu > 0 > \int \xi \d \mu^* \ge \min_{\mu \in D} \int \xi \d \mu,
\end{align*}
from which $\xi \in \calU(0)$ but $\xi \notin \calV(0)$, a contradiction.

Lastly, let us show that $\underline{T}(\xi) = \min_{\mu \in C} \int \xi \d \mu$ and $\overline{T}(\xi) = \min_{\mu \in D} \int \xi \d \mu$ for all $\xi \in \R^\Omega$.
Suppose not, $\underline{T}(\xi) > \min_{\mu \in C} \int \xi \d \mu$ holds for some $\xi$.
By (\ref{claim_add2}), we can find $r \in \R$ such that $\underline{T}(\xi + r\1_\Omega) > 0 > \min_{\mu \in C} \int (\xi+r\1_\Omega) \d \mu$.
But then, (\ref{contour_maxmin1}) implies $\xi + r\1_\Omega \in \calU(0)$, while (\ref{contour_maxmin2}) implies $\xi+r\1_\Omega \notin \calU(0)$, a contradiction.
Similarly, we would encounter a contradiction if $\underline{T}(\xi) < \min_{\mu \in C} \int \xi \d \mu$.
Thus, $\underline{T}(\xi) = \min_{\mu \in C} \int \xi \d \mu$ must hold.
Analogously, we can show that $\overline{T}(\xi) = \min_{\mu \in D} \int \xi \d \mu$.
Together with the representation (\ref{eq_pf_maxmin}), we obtain the desired twofold maxmin representation of $\succ$.

Clearly, $u$ is unique up to positive affine transformations, and the uniqueness of $(C,D)$ follows from the standard separating hyperplane arguments.
\hfill {\it Q.E.D.}

\subsubsection{An Example of Completion that Violates (R1)} \label{appendix_example}

We argue that (R1) is indispensable for Proposition \ref{prop_completion}, in the sense that there exists a completion $\succ^*$ of $\succ$ that violates (R1) and cannot be represented in a generalized $\alpha$-maximin form.
Let $\succ$ admit a symmetric twofold multiprior representation $(u,C,C)$.
Assume that $|C| \neq 1$, $[-1,1] \subseteq X \subseteq \R$, and that $u(x) = x$ for all $x \in X$.
Consider the following utility functions:
\begin{align*}
I(f) &= (1-\alpha) \min_{\mu \in C} \int u(f) \d\mu + \alpha \max_{\mu \in C} \int u(f) \d\mu, \\
J(f) &= (1-\beta) \min_{\mu \in C} \int u(f) \d\mu + \beta \max_{\mu \in C} \int u(f) \d\mu,
\end{align*}
where $\alpha$ and $\beta$ are fixed numbers such that $0 \le \alpha < \beta \le 1$.
By Proposition \ref{prop_completion}, each preferences represented by $I$ or $J$ becomes a regular completion of $\succ$.

Beside these completions, we consider an asymmetric preference $\succ^*$ defined by the following lexicographic rule:
\begin{align*}
f \succ^* g \Longleftrightarrow
\begin{cases}
I(f) > I(g); \text{ or} \\
I(f) = I(g) \text{ and } J(f) > J(g).
\end{cases}
\end{align*}
It is not hard to see that $\succ^*$ is a completion of $\succ$ and satisfies (R2).
We claim, however, that $\succ^*$ violates (R1).
To this end, since $|C| \neq 1$ and $[-1,1] \subseteq X$, we can take $f \in \calF$ and $\underline{x}, \overline{x} \in X$ such that 
\[
\underline{x} < \min_{\mu \in C} \int u(f) \d\mu < \max_{\mu \in C} \int u(f) \d\mu < \overline{x}.
\]
By the definition of $\succ^*$, for any $\lambda \in [0,1]$, it follows that
\begin{align*}
&f \succ^* (1-\lambda) \underline{x} + \lambda \overline{x} \\
&\Longleftrightarrow \bigl[ I(f) > (1-\lambda) \underline{x} + \lambda \overline{x} \bigr] \ \text{ or } \ \bigl[ I(f) = (1-\lambda) \underline{x} + \lambda \overline{x} \ \text{ and } \ J(f) > (1-\lambda) \underline{x} + \lambda \overline{x} \bigr] \\
&\Longleftrightarrow \bigl[ I(f) > (1-\lambda) \underline{x} + \lambda \overline{x} \bigr] \ \text{ or } \ \bigl[ I(f) = (1-\lambda) \underline{x} + \lambda \overline{x} \bigr] \\
&\Longleftrightarrow \bigl[ I(f) \ge (1-\lambda) \underline{x} + \lambda \overline{x} \bigr],
\end{align*}
where the second equivalence is due to the fact that $I(f) < J(f)$, which follows from $\alpha < \beta$.
The above argument shows that $\{\lambda \in [0,1]: f \succ^* (1-\lambda) \underline{x} + \lambda \overline{x}\}$ is not open relative to $[0,1]$, so (R1) fails.

\newpage

\bibliography{main}

\end{document}